\documentclass[aps, prb, preprint]{revtex4}

\usepackage{graphicx}
\usepackage{amsmath}
\usepackage{amssymb}
\begin{document}
\pacs{xx.yy.MM, 00.00.NN}
 \title{Wave-function-based approach to quasiparticle bands: new insight into the electronic structure of $c$-ZnS}
 
\author{A.\ Stoyanova$^{1}$}
\email[Corresponding authtor:]{alex07@mpipks-dresden.mpg.de; stoyanova@unistra.fr}
\author{L.\ Hozoi$^{1, 2}$}
\author{P.\ Fulde$^{1, 3}$}

\affiliation{$^{1}$Max-Planck-Institut f\"ur Physik komplexer Systeme, \\
                   N\"othnitzer Strasse 38, 01187 Dresden, Germany \\
                   $^{*}$ present address: Laboratoire de Chimie Quanitique, Institut de Chimie, CNRS/           Universit\'e de Strasbourg, \\
                   4 rue Blaise Pascal, 6700 Strasbourg, France}
\affiliation{$^{2}$Institut f\"ur Theoretische Festk\"orperphysik, \\
Leibniz-Institut f\"ur  Festk\"orper- und Werkstofforschung, \\
Helmholzstr. 20, 01069 Dresden, Germany
                }   
 \affiliation{$^{3}$Asia Pacific Center for Theoretical Physics \\
Hogil Kim Memorial Building 501 \\
POSTECH, San 31 Hyoja-dong, Namgu \\
Pohang, Gyeongbuk 790-784, Korea  \\
                }                                
\author{H. Stoll$^{4}$}
\affiliation{$^{4}$Institut f\"ur Theoretische Chemie, Universit\"at Stuttgart, \\ Pfaffenwaldring 55, D-70569 
Stuttgart, Germany                 
               }
\begin{abstract}
$Ab$ $initio$ wave-function-based methods are employed for the study of quasiparticle energy bands of zinc-blende ZnS, with focus on the Zn $3d$ ``semicore"  states. The relative energies of these states with respect to the top of the S $3p$ valence bands appear to be poorly described as compared to experimental values not only within the local density approximation (LDA), but also when many-body corrections within the $GW$ approximation are applied to the LDA or LDA+$U$ mean-field solutions [T. Miyake, P. Zhang, M. L. Cohen, and S. G. Louie, Phys. Rev. B \textbf{74}, 245213 (2006)]. 
In the present study, we show that for the accurate description of the Zn $3d$ states a correlation treatment based on wave function methods is needed. 
Our study rests on a local Hamiltonian approach which rigorously describes the short-range polarization and charge redistribution effects around an extra hole or electron placed into the valence respective conduction bands of semiconductors and insulators. 
The method also facilitates the computation of electron correlation effects beyond relaxation and polarization. 
The electron correlation treatment is performed on finite clusters cut off the infinite system.
The formalism makes use of localized Wannier functions and embedding potentials derived explicitly from prior periodic Hartree-Fock calculations. The on-site and nearest-neighbor charge relaxation lead to corrections of several eV to the Hartree-Fock band energies and gap. Corrections due to long-range polarization are of the order of 1.0 eV. The dispersion of the Hartree-Fock bands is only little affected by electron correlations. 
We find the Zn $3d$ ``semicore"  states to lie about 9.0 eV below the top of the S $3p$ valence bands, in very good agreement with values from valence-band x-ray photoemission. 

 \end{abstract}
\maketitle
\section{Introduction}
The accurate \textit{ab initio} treatment of many-body effects in the ground and excited states of solids remains a challenging problem in physics. 
For the ground-state (GS) properties of periodic systems reliable \textit{ab initio} methods are at present available. Two different formalisms can be distinguished: approaches based on Kohn-Sham (KS) density functional theory (DFT) \cite{KS} and 
methods which rest on explicitly computing many-body wave functions (WF's) \cite{Fuldebook, Fulde}, by means of quantum chemical techniques \cite{Helgaker}. The accuracy achieved by the two types of methods for the GS properties is often comparable. For the case of energy bands and excited states, however, the results obtained by KS-DFT and WF-based approaches are different. 
Although DFT formally is a GS theory, it has been commonly employed in the computation of energy bands, i.e., excited states, and in some cases yields surprisingly good results, e.g., when hybrid density functionals or exact exchange schemes are utilized. Nevertheless, a 
rigorous justification for assigning the KS eigenvalues or KS energy differences to excitation or quasiparticle energies of correlated electron systems is still to be given. 
Since there is no solid basis for such an extension of GS KS-DFT, when inaccurate or wrong excitation energies are obtained, they can not be improved in a systematic and controlled manner. 

The widely used local density (LDA) \cite {Perdew} and generalized gradient approximations (GGA) \cite{Perdew2} to the KS-DFT provide a rather poor description for binding energies and band gaps of semiconductors and insulators.
The self-interaction error introduced with the LDA or GGA potential and the absence of discontinuity in the exchange potential have been identified as the main reasons for that. 
The incomplete cancellation of the self-interaction in LDA or GGA is also regarded as a major cause for the underestimation of the binding energies of ``semicore" $d$ states in systems such as the II$^B$-VI$^{A}$ semiconductors \cite{Miyake, Sharma, Zhang}. An accurate \textit{ab initio} description of the electronic structure of these seemingly conventional $sp$ semiconductors has appeared to be a rather difficult task.

Numerous refinements of KS-DFT have been suggested to enable the treatment of many-body effects in correlated electron systems. 
Many-body corrections to the KS mean-field eigenvalues as obtained within the $GW$ approximation \cite{Hedin} to the self-energy often yield good agreement between theory and experiment with respect to the band gaps of various group-IV and III-V semiconductors. The $GW$ approximation is established as the standard method for calculating quasiparticle energy bands in those materials.  
It does fail, however, to describe complex many-body effects in $d$-electron systems such as the II$^B$-VI$^{A}$ compounds \cite{Miyake, Luo, Fleszar, Rohlfing, Zakharov}. The quasiparticle energies of the $d$ states, for example, are about 2.0 eV higher than the experimental values \cite{dpos1, dpos2}. This result remains the same when the mean-field LDA+\textit{U} \cite{Anisimov} Hamiltonian is utilized \cite{Miyake} as a basis for the $GW$ treatment \footnote{The final $GW$ quasiparticle energies are supposed to be independent of the starting mean-field solution.}. 
The problems were attributed to the self-interaction errors, still present in the standard $GW$ approximation \cite{Fuchs}. 

It has been argued that determining self-consistently the optimum mean-field Hamiltonian rather than utilizing the LDA (+\textit{U}) would largely improve the description of quasiparticles within the $GW$ approximation \cite{Kotani1}. Indeed, the self-consistent $GW$ method of Kotani \textit{et al.} \cite{Kotani1} yields in the II$^B$-VI$^{A}$ and III$^B$-V$^{A}$ compounds $d$-state quasiparticle energies in good agreement with experimental values, i.e., deviations of at most 0.8 eV are observed. Yet, the approach overestimates systematically, although by little, the band gaps. This was attributed to the lack of excitonic effects in the dielectric function.

Methods which treat the exchange energy within the exact exchange (EXX) scheme and are thus considered to be self-interaction free have been developed as well. 
For example, the full-potential EXX KS-DFT calculations of Sharma \textit{et al.} \cite{Sharma} including core-valence exchange interactions yield very good agreement with the experiment for the binding energies of $d$ electrons in ZnS. 
The band gaps of the II$^B$-VI$^{A}$ compounds and other $sp$ semiconductors, such as Si, Ge, and GaAs,  are, however, largely overestimated in that computational scheme. This trend is reversed in insulating compounds such as BN, C and some nobel gas solids, where the gaps are underestimated \cite{Sharma}. Excluding the core-valence exchange interaction, on the other hand, brings the band gaps close to experimental estimates, but it simultaneously worsens the binding energies. EXX calculations within the pseudopotential framework yield band gaps and cohesive properties of various $sp$ semiconductors in very good agreement with the experiment \cite{Stadele}. The binding energies of the $d$ states, however, remain wrong in such a treatment \cite{Rinke}.  

A different approach to energy bands in periodic systems is based on \textit{ab initio} WF methods. Their major virtue is the use of well-defined and controllable approximations.
Converged results for binding energies, lattice parameters, and band gaps can be here achieved by systematic improvement of the many-body wave function \cite{Fulde}. 
The disadvantage of those formalisms is the larger amount of work involved as compared to the DFT-based calculations. 
The large computational effort is related to the physics of the problem. When an electron is added to a conduction-band or removed from a valence-band state of a semiconductor or insulator, the correlation hole of the extra particle (hole) is long-ranged because of polarization and relaxation effects. The corresponding  quasiparticle, i.e., the particle plus its correlation hole, is moving in form of a Bloch wave through the system. The computation of the energy dispersion associated with the quasiparticle 
requires an accurate description of the correlation hole. The latter differs considerably from the correlation hole of an electron in the ground state for which the correlations are of van der Waals type and do not involve long-range (LR) polarizations.
The different nature of the correlation holes in the ground- and ($N\!\pm\!1\!$) states is simply ignored in the DFT methods and hence the large computational effort is avoided. Taking into account this difference  
is essential for the correct determination of energy bands and gaps. 

In contrast to ground-state calculations \cite{Karin, Hirata2, Ladik,  Hirata, Stoll1, Shukla2, Shukla1, Paulus}, energy band calculations based on quantum chemical methods are still limited to non-metalic systems \cite{Sun, Ayala, Pisani, Malrieu, Forner, Stoll1, Shukla2, Birken2, Gr1, Albrecht, Birken1, GF2, Shukla1, Abdurahman, Pisani2008, Stollhoff}. 
They rest either on second-order M\o ller-Plesset theory (MP2) \cite{Sun, Ayala, Pisani, Malrieu}, coupled-cluster (CC) techniques \cite{Forner} or effective local Hamiltonian (LH) \cite{Horsch1, Gr1, Birken1, Birken2, Albrecht, BezuglyJP} approaches. 
Some of these formalisms make use of the local character of the short-range part of the correlation hole \cite{Gr1, Albrecht, Birken1, Birken2, GF2, Pisani, Malrieu, Shukla1, Shukla2, Pisani2008, Stollhoff, Abdurahman}, utilizing thereby local operators \cite{Fuldebook, Fulde} related to sets of real-space, localized orbitals. 
The localized orbitals are either derived by means of various localization procedures \cite{Foster-Boys, Edmiston, Pulaypaos} applied to the canonical Bloch orbitals of the self-consistent-field (SCF) GS wave function \cite{Pisani, Zikovich, disent, Marzari, Forner, Karin} or obtained directly from a crystal Hartree-Fock \cite{HF} calculation carried out in the Wannier representation \cite{Shukla2, Shukla1, Malrieu}. 
The local orbital schemes of Pulay and Saeb$\o$ utilized in local M\o ller-Plesset and coupled-electron-pair correlation approaches to finite systems \cite{Pulay1, Pulay2} have also been adopted for deriving local orbitals for crystalline compounds, see, e.g., Ref. ~\onlinecite{Birken2}. 

In the present study, we compute the correlation-induced corrections to the HF valence and conduction energy bands of $c$-ZnS, a representative of the 
the II$^B$-VI$^{A}$ compounds.
There is much interest toward the electronic structure and properties of the II$^B$-VI$^{A}$ systems due to their optical properties \cite {Exper1, Exper2, Exper3, Exper4} and potential exploitation in optoelectronic devices and medical instrumentation. For example, thin films of these materials grown on substrates of GaAs give rise to interfaces emitting intensive white light, i.e., semiconductor-based light emitting devices \cite{TP1, TP2}.  
Moreover, some properties of mixed phases of the type ZnS$_{x}$Se$_{1-x}$  such as the width of the fundamental band gap may be continuously  tuned depending on the explicit $x$ value \cite{Homann}.

The goal of these WF-based investigations is twofold. On one hand, we extend the area of applicability 
of the WF-based band structure calculations, building on previous investigations performed for the simpler compounds MgO \cite{Hozoi} and  BN \cite{AlexJCP}. On the other hand, 
our study attempts to provide new insight into the many-body effects which operate in this specific $3d$ electron system. 
Of particular interest are the binding energies of the ``semicore" Zn 3$d$ states since their accurate treatment remains out of reach for standard DFT and $GW$ methods, see Refs. ~\onlinecite{Miyake}, ~\onlinecite {Luo} and the discussion above. 
HF calculations carried out within the linear-combination-of-atomic orbital method \cite{Jaffe2} or the full-potential linear-muffin-tin-orbitals approach \cite{Kotani2} overestimate as well, by about 3-5 eV, the binding energies of $d$ states in II$^B$-VI$^A$ and III$^A$-V$^A$ semiconductors.

Our WF-based studies are performed in the framework of the LH approach \cite{Horsch1, Albrecht, Gr1, Birken2, Birken1} and quasiparticle approximation \cite{Fulde}, where the quasiparticle energy bands are expressed in terms of matrix elements of an effective local Hamiltonian. A short overview of the method is presented in Section II, since a detailed account has already been published \cite{Birken1, Gr1, Fulde, Hozoi,  AlexJCP}. 

\section{Theoretical framework}
\subsubsection{Quasiparticle energy bands}
The \textit{correlated} quasiparticle valence- and conduction-band energies of periodic (closed-shell) systems are defined as \cite{Fuldebook, Gr1, Birken2, Albrecht, Birken1, BezuglyJP}
\begin{eqnarray}
\epsilon_{\mathbf{k}\mu\sigma}&=&E^{N}_{0}-\underbrace{\langle \Psi_{\mathbf{k}\mu\sigma}^{N-1}|H|\Psi_{\mathbf{k}\mu\sigma}^{N-1}\rangle}_{E^{N-1}_{\mathbf{k}\mu\sigma}},  
\nonumber \\
\epsilon_{\mathbf{k}\nu\sigma}&=& \underbrace{\langle \Psi_{\mathbf{k}\nu\sigma}^{N+1}|H|\Psi_{\mathbf{k}\nu\sigma}^{N+1}\rangle}_{E^{N+1}_{\mathbf{k}\nu\sigma}}-E^{N}_{0},
\label{epsiloncorr}
\end{eqnarray}
where $E^{N}_{0}$ is the energy of the correlated ground state of the neutral $N$-electron crystal and $E^{N+1}_{\mathbf{k}\nu\sigma}$ and $E^{N-1}_{\mathbf{k}\mu\sigma}$ are the correlated energies of the crystal ($N\!+\!1\!$) and ($N\!-\!1\!$) electron states, respectively. 
The quasiparticle states $|\Psi_{\mathbf{k}\nu\sigma}^{N+1}\rangle$ of the ($N\!+\!1\!$) electron system are related to the corresponding SCF counterparts $|\Phi^{N+1}_{\mathbf{k}\nu\sigma}\rangle$ through \cite{Birken2, Fuldebook, Fulde}
\begin{eqnarray}
|\Psi_{\mathbf{k}\nu\sigma}^{N+1}\rangle&=&\Omega|\Phi^{N+1}_{\mathbf{k}\nu\sigma}\rangle \label{WT1},
\end{eqnarray}
where $\nu$ is a band index, and $\sigma$ and $\mathbf{k}$ are the spin and momentum of the ($N\!+\!1\!$) state. The wave operator $\Omega$ in a coupled cluster formulation is given by $\Omega=e^{S}$, where $S$ is a scattering operator.
A similar relation holds for the ($N\!-\!1\!$) states $|\Psi^{N-1}_{\mathbf{k}\mu\sigma}\rangle$ of momentum $\mathbf{-k}$, band index $\mu$, and spin $-\sigma$ obtained when an electron is removed from the system.  

The SCF ($N\!+\!1\!$) states are expressed in terms of local electron-addition single-particle configurations $|\Phi^{N+1}_{\mathbf{R}_{I}n\sigma}\rangle$ \cite{Gr1, Birken1, Birken2, BezuglyJP} 
\begin{eqnarray}
|\Phi^{N+1}_{\mathbf{k}\nu\sigma}\rangle&=&\frac{1}{\sqrt{N_{0}}}\sum_{n, \mathbf{R}_{I}}\alpha_{\nu n}(\mathbf{k})e^{i\mathbf{k}.\mathbf{R}_{I}} \underbrace{w^{\dagger}_{\mathbf{R}_{I}n\sigma} |\Phi_{\textsc{scf}}\rangle}_{|\Phi^{N+1}_{\mathbf{R}_{I}n\sigma}\rangle}, 
\label{WT2}
\end{eqnarray}
where $N_{0}$ is the total number of unit cells and $\alpha_{\nu n}(\mathbf{k})$ is the structure matrix \cite{Fulde}. The local operator $w^{\dagger}_{\mathbf{R}_{I}n\sigma}$ creates an electron in a real-space localized conduction-band Wannier orbital (WO) $|w_{n\sigma} (\mathbf{R}_{I})\rangle$ of index $n$, centered at a site defined by the lattice vector $\mathbf{R}_{I}$ in unit cell \textit{I}. 
The SCF GS wave function $|\Phi_{\textsc{scf}}\rangle$ and energy $E_{0}^{\textsc{scf}}$ of the $N$-electron system as well as the WO's are derived via a periodic HF calculation for the extended system. 
Analogously to the HF ($N\!+\!1\!$) states, the HF states $|\Phi^{N-1}_{\mathbf{k}\mu\sigma}\rangle$ are expressed in terms of local single-hole configurations $|\Phi^{N-1}_{\mathbf{R}_{I}m\sigma}\rangle$ by utilizing local annihilation operators $w_{\mathbf{R}_{I}m\sigma}$ for the valence-band WO's $|w_{m\sigma} (\mathbf{R}_{I})\rangle$. 
By combining Eqs. (\ref{WT1}) and (\ref{WT2}) with Eq. (\ref{epsiloncorr}), the correlated conduction-band energies are expressed in terms of the matrix elements of the local effective Hamiltonian $H^{eff}$ %
\begin{eqnarray}
\epsilon_{\mathbf{k}\nu\sigma}&=&
\sum_{\mathbf{R}_{I}}\sum_{nn'}\alpha_{\nu n}(\mathbf{k})\alpha^{*}_{\nu n'}(\mathbf{k}) e^{i\mathbf{k}.\mathbf{R}_{I}} \times
\nonumber \\
&\times& 
\underbrace{\langle\Phi_{\textsc{scf}}|w_{\mathbf{0}n'\sigma}\Omega^{\dagger}H \Omega w^{\dagger}_{\mathbf{R}_{I}n\sigma}|\Phi_{\textsc{scf}} \rangle }_{H^{eff}_{\mathbf{0}n'\sigma,\mathbf{R}_{I}n\sigma}}
\nonumber \\
&-&\underbrace{E^{N}_{0} \delta_{\mathbf{R}_{I}\mathbf{0}}\delta_{nn'}}_{H^{eff}_{\mathbf{0}n'\sigma,\mathbf{R}_{I}n\sigma}},
\label{epsiloncorr2}
\end{eqnarray}
where $E^{N}_{0}=\langle\Phi_{\textsc{scf}}|\Omega_{0}^{\dagger}H\Omega_{0}|\Phi_{\textsc{scf}}\rangle$ and the wave operator $\Omega_{0}$ transforms $\Phi_{\textsc{scf}}$ into the true ground state $\Psi^{N}_{0}$ of the $N$-particle system. 
In Eq. \ref{epsiloncorr2}, the matrix elements of $H^{eff}$    
are defined in terms of the real-space \textit{correlated} ($N\!+\!1\!$) electron states $|\Psi_{\mathbf{R}_{I}n\sigma}^{N+1}\rangle$=$\Omega w^{\dagger}_{\mathbf{R}_{I}n\sigma}|\Phi_{\textsc{scf}}\rangle$. A Wannier-like transformation relates $|\Psi_{\mathbf{R}_{I}n\sigma}^{N+1}\rangle$ with their $\mathbf{k}$-space counterparts $|\Psi_{\mathbf{k}\nu\sigma}^{N+1}\rangle$. 

To determine the correlation corrections to the SCF band energies, $\epsilon^{corr}_{\mathbf{k}\nu\sigma}$=$\epsilon_{\mathbf{k}\nu\sigma}$--$\epsilon^{\textsc{scf}}_{\mathbf{k}\nu\sigma}$, 
the SCF energies $\epsilon^{\textsc{scf}}_{\mathbf{k}\nu\sigma}$ are first computed by utilizing the local matrix elements 
$H^{\textsc{scf}}_{\mathbf{0}n',\mathbf{R}_{I}n}$=$\langle\Phi_{\textsc{scf}}|w_{\mathbf{0}n'\sigma}Hw^{\dagger}_{\mathbf{R}_{I}n\sigma}|\Phi_{\textsc{scf}}\rangle$--$E_{0}^{\textsc{scf}} \delta_{\mathbf{R}_{I}\mathbf{0}\delta_{nn'}}$. The expressions for the valence-band energies, corresponding matrix elements and correlated wave functions are analogous. 

Because of the predominantly local character of the correlation hole, it suffices to compute all relevant matrix elements $H^{eff}_{\mathbf{0}n',\mathbf{R}_{I}n}$ and $H^{eff}_{\mathbf{0}m',\mathbf{R}_{I}m}$ using sufficiently large clusters cut off the infinite system \cite{Gr1, Birken1, Birken2, Hozoi, AlexJCP}. 
These matrix elements can be determined by utilizing CC or multiconfiguration (MC) and multireference (MR) configuration-interaction (CI) calculations \cite{Helgaker} when correlations are strong. In principle, the correlation treatment can be facilitated by the method of increments, where the total correlation energy is assembled from contributions originating from selected groups of localized orbitals \cite{Stoll1, Stoll2, Paulus}.

The scattering operator \textit{S} is given by a selected set of excitation operators, see, e. g., Refs.\ ~\onlinecite{Fuldebook},\ ~\onlinecite{Fulde}.  
For relatively weakly correlated systems, a reasonable ansatz for \textit{S} includes one- and two-particle excitations which generate 
the correlation hole of the added particle, i.e., the relaxation and polarization cloud, and account for the loss of ground-state correlations \cite{Gr1, Fuldebook, Birken2, Fulde}.
\subsubsection{Computation of the matrix elements of H$^{eff}$}
To obtain the correlation corrections to the matrix elements of H$^{eff}$,
$H_{\mathbf{0}n',\mathbf{R}_{I}n}-H^{\textsc{scf}}_{\mathbf{0}n',\mathbf{R}_{I}n}$, we calculate the short- and long-range part of the correlation hole of an added electron or hole. The short-range (SR) part of the correlation hole consists of intra-atomic and SR inter-atomic relaxation and polarization. 
This response of the nearest surroundings is determined by means of SCF orbital optimizations for a  
finite cluster around the extra electron or hole set into a conduction-band or valence-band WO.   
The WO hosting the added particle is kept frozen in the SCF calculation \cite{MCSCF} for the ($N\!\mp\!1\!$) excited state, which is referred to as the frozen local hole (electron) approximation \cite{Birken2, Pahl}. 
In first-order perturbation theory, the SCF orbital relaxation corresponds to single-particle excitations around the frozen WO, see, Ref. ~\onlinecite{Pahl}. The corresponding wave functions $|\tilde{\Psi}_{\mathbf{R}_{I}m\sigma}^{N-1}\rangle$ and $|\tilde{\Psi}_{\mathbf{R}_{I}n\sigma}^{N+1}\rangle$ \footnote{The wave functions $|\tilde{\Psi}_{\mathbf{R}_{I}m\sigma}^{N-1}\rangle$ and $|\tilde{\Psi}_{\mathbf{R}_{I}n\sigma}^{N+1}\rangle$ are correlated only through single-particle excitations.} are thus given by the optimized SCF states $|\tilde{\Phi}_{\mathbf{R}_{I}m\sigma}^{N-1}\rangle$ and $|\tilde{\Phi}_{\mathbf{R}_{I}n\sigma}^{N+1}\rangle$, respectively. 
For the case of weakly and moderately correlated systems, the SCF calculation for the finite region suffices to describe the SR correlation hole. 
Because the local operators are associated with a set of \textit{spin orbitals}, while, in practice, a \textit{spatial orbital} set is utilized, the spin degree of freedom is accounted for by further optimizing the WO with the extra particle within the previously relaxed environment, see Ref. ~\onlinecite{AlexJCP}. The wave functions expressed in terms of relaxed orbitals within the nearby surroundings, including the reoptimized singly-occupied WO, are denoted as $|\breve{\Phi}_{\mathbf{R}_{I}m\sigma}^{N-1}\rangle$ and $|\breve{\Phi}_{\mathbf{R}_{I}n\sigma}^{N+1}\rangle$. 

In distinction to the correlation hole of an electron in the GS of a system, which extends only over few lattice sites \cite{Fuldebook, Fulde, Hozoi2}, the correlation hole of an extra particle in the ($N\!\pm\!1\!$) system includes a substantial long-range (LR) polarization cloud. This effect originates from the polarization of the farther surroundings and can not be described by calculations on finite clusters.  
Yet, only the SR part of the correlation hole is essential when the energies of different ($N\!+\!1\!$) or ($N\!-\!1\!$) states are compared because the contribution of the LR polarization cloud cancels out. The LR polarization cloud is treated here within a dielectric continuum approximation using the experimental dielectric constant \cite{Gr1, Fuldebook, Birken2, Hozoi, AlexJCP}.  Its contribution to the correlation-induced corrections to the matrix elements of $H^{eff}$ is explicitly incorporated and hence ionization potentials and electron affinities can be evaluated as well.

Correlation effects beyond charge relaxation and polarization, including the loss of ground-state correlations, are computed by means of single and double CI (SDCI) calculations \cite{ICCI} on top of the SCF orbital relaxation, separately for each $|\tilde{\Phi}_{\mathbf{R}_{I}m\sigma}^{N-1}\rangle$ and $|\tilde{\Phi}_{\mathbf{R}_{I}n\sigma}^{N+1}\rangle$ or $|\breve{\Phi}_{\mathbf{R}_{I}m\sigma}^{N-1}\rangle$ and $|\breve{\Phi}_{\mathbf{R}_{I}n\sigma}^{N+1}\rangle$ state. 
The loss of ground-state correlations is a measure of correlation contributions that are present in the $N$-electron ground state but absent in the ($N\pm1$) states, since in the latter some excitations are blocked.  
To account for such effects, the conduction-band WO's to which an extra electron is attached and the valence-band WO's out of which an electron is removed are always kept singly occupied in the SDCI wave functions.
To treat in a unified manner both valence- and conduction-band states and to avoid spurious charge flows within the finite cluster $\mathcal{C}$, the \textit{infinite} frozen crystal environment of $\mathcal{C}$ is explicitly incorporated in the calculations. 
The effect of the electrons in the localized occupied orbitals of the environment $\mathcal{E}$ on the electrons in the localized orbitals of $\mathcal{C}$ is subsumed in an effective one-electron potential $V^{emb}$, see, e.g., Refs. ~\onlinecite{McWeeny}, ~\onlinecite{Werner}. 
The embedding potential $V^{emb}$ is computed 
by utilizing the crystal self-consistent Fock operator $F^{cryst}$ and the Fock operator of $\mathcal{C}$, $F[P^{\mathcal{C}}]$, associated with the density operator $P^{\mathcal{C}}$ \cite{Birken2, Hozoi, AlexJCP}. In the cluster calculations, $V^{emb}$=$F^{cryst}-F[P^{\mathcal{C}}]$ is added back to $F[P^{\mathcal{C}}]$ and the correlation calculations are thus effectively performed in an infinite frozen HF environment. 

Our embedded cluster formalism requires Wannier-Boys localization \cite{Zikovich} of the HF core, valence, and lowest-lying conduction-band states. The optimally localized valence-band and conduction-band WO's which are centered at lattice sites within $\mathcal{C}$ are then projected onto the set of Gaussian-type-orbital (GTO) basis functions assigned to those cluster sites \cite {Birken2, Hozoi, AlexJCP}. Their longer-range tails that involve Gaussian basis functions at sites outside $\mathcal{C}$ and orthogonal to the basis functions in $\mathcal{C}$ are thus cut off. 
Hence, the projected WO's that are centered at sites within the central part of $\mathcal{C}$ bear closest resemblance to the original WO's, while those near the edges of $\mathcal{C}$ are represented asymmetrically in the basis set of $\mathcal{C}$. For the post-HF treatment, the cluster $\mathcal{C}$ is thus subdivided into a central active region $\mathcal{C}_A$, and a buffer region $\mathcal{C}_B$. The role of the sites and corresponding basis functions within $\mathcal{C}_B$ is to ensure an accurate representation of the longer-range tails of the WO's centered within $\mathcal{C}_A$. For sufficiently large $\mathcal{C}_B$ regions, the original WO's in $\mathcal{C}_A$ remain in practice unaffected by the projection, i.e., their projected counterparts  
show the same properties as the original ones \cite{Birken2, Hozoi, AlexJCP}. 
In the post-HF calculation, the occupied and lowest-lying conduction-band WO's centered within $\mathcal{C}_A$ are explicitly correlated while the WO's within $\mathcal{C}_B$ are held frozen.

The projected WO's are neither normalized nor orthogonal. Thus, they are groupwise orthonormalized in the order active core, active valence, active lowest-lying conduction-band orbital groups, buffer core, buffer valence, and buffer lowest-lying conduction-band orbitals \cite{Birken2, Hozoi, AlexJCP}. Our localized ($N\pm1$) states are expressed in terms of those projected WO's.

To generate the virtual orbital space, the projected atomic orbital (PAO) formalism of Pulay \cite{Pulaypaos} and further elaborated by Hampel and Werner \cite{Hampel} is adopted. 
The virtual PAO's are constructed from the GTO basis functions assigned to the sites in $\mathcal{C}_A$ after projecting out the orthonormalized occupied and lowest-lying conduction-band WO's of $\mathcal{C}$ via a Schmidt orthogonalization, see also Refs. ~\onlinecite{Hozoi} and ~\onlinecite{AlexJCP}. They are subsequently L\"owdin orthonormalized among themselves. This procedure yields thus PAO's orthogonal to those occupied and lowest-lying conduction-band WO's. While the PAO's in $\mathcal{C}_A$ are identical with the virtual orbitals of the crystal, the PAO's in $\mathcal{C}_B$ deviate from those of the infinite system. In the correlation calculations, however, the variational virtual orbital space consists only of PAO's within $\mathcal{C}_A$.
\section{Crystal structure of zinc-blende ZnS and computational information}
ZnS is representative of the II$^B$-VI$^A$ semiconductors crystallizing at ambient pressure in the cubic zinc-blende structure.
The zinc-blende form is face-centered (fcc), with the $F\bar{4}3m$ space group symmetry. The corresponding lattice constant is 5.415 \AA \, \cite{Lattconst}.
The local point-group symmetry for both Zn and S is $T_{d}$, with a four site nearest-neighbor
coordination. 

GTO basis sets of triple-zeta quality augmented with polarization functions are applied in our study for
the S atoms. We employed the 86-311G* GTO basis set of Lichanot \textit{et al.} \cite{Lichanot} derived for crystalline MgS but reoptimized in periodic HF calculations the exponents of the outermost two $sp$ shells as well as that of the single-Gaussian $d$ polarization function. The exponents of the reoptimized $sp$ shells are $0.27202$ and
$0.125$. The reoptimized $d$ polarization function has an exponent of $0.435$. The latter value is very close to the one of 0.479 used in correlated molecular calculations \cite{Dunning}.

The basis set used for Zn is a 86-4111d311 GTO basis set which is derived from the 86-4111d41G basis set constructed for ZnO by Jaffe \textit{et al.} \cite{Jaffe} and recently employed for ZnS$_{x}$Se$_{1-x}$ by Homann \textit{et al.} \cite{Homann}. 
The basis set of Jaffe \textit{et al.} was properly modified to achieve a triple-zeta basis for the description of the $d$ states. For this purpose, the most diffuse primitive of the $3d$ shell was decontracted and the exponents and contraction coefficients of the valence $3d$ shell as well as the exponents of the outermost single-Gaussian polarization $4sp$, $5sp$, $6sp$ and $4d$, $5d$ functions were reoptimized. Their optimized values are presented in Table \ref{BS1}. 
The exponents of the outer $sp$ and $d$ shells underwent almost no variations during the HF optimization as compared to those reported by Jaffe \textit{et al.}. For the Zn $3d$ states, this outcome is due to their ``semicore" nature.

The periodic HF calculations were carried out with the \textsc{crystal} program \cite{CRYSTAL1}. The Wannier-Boys orbital localization module in \textsc{crystal} was employed to construct localized Wannier functions \cite{Zikovich}. 
The projected WO's for the finite cluster $\mathcal{C}$ 
and the matrix representation $F^{cryst}_{\alpha\beta}$ of the crystal
Fock operator in the basis set of $\mathcal{C}$
are both obtained with the \textsc{crystal-molpro} interface program \cite{interface}. In addition, the \textsc{crystal-molpro} interface generates the virtual PAO's. The Fock operator of $\mathcal{C}$, $F[P^{\mathcal{C}}]$, and the density operator $P^{\mathcal{C}}$ are calculated with the \textsc{molpro}  \cite{molpro} suite of programs.
The embedding potential for $\mathcal{C}$ is also computed with the help of \textsc{molpro}. 
The post-HF correlation calculations are carried out with the \textsc{molpro} package as well.
\section{Correlation effects on the band structure of zinc-blende ZnS}
\subsection{HF band structure}
\begin{figure}[htbp]
\begin{center}
\includegraphics[width=8.5cm]{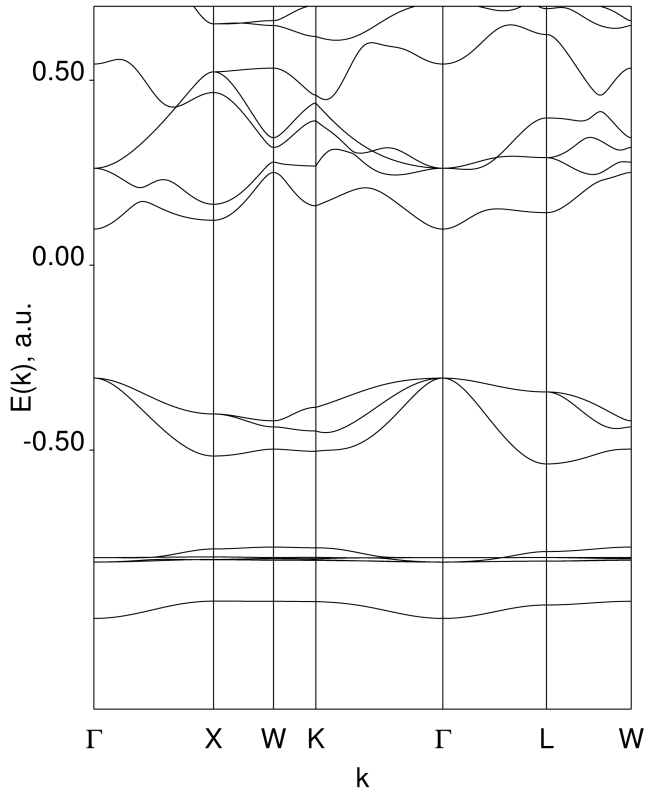}
\caption{Hartree-Fock band structure of ZnS.}
\label{hfband}
\end{center}
\end{figure}

\begin{figure}[htbp]
\begin{center}
\includegraphics[width=8.5cm]{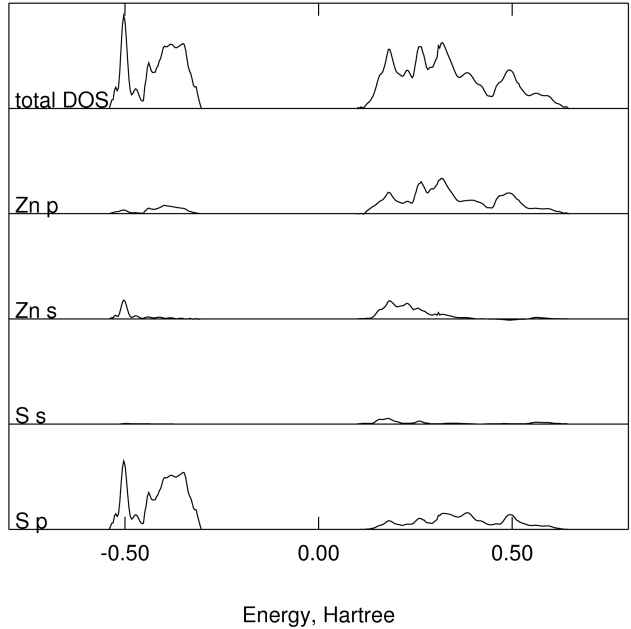}
\caption{Total and projected density of states (DOS) of ZnS around the Fermi level.}
\label{dos}
\end{center}
\end{figure}

The HF band structure of ZnS, computed with the triple-zeta GTO basis sets described in Sec. III, is shown in Fig.\ \ref{hfband}. 
The upper valence bands have dominant S $3p$ character with very small contributions from the Zn $4s$ and $4p$ states. 
The projected density of states (DOS) indicate that the lowest conduction bands have predominantly Zn $4s$ and $4p$ character, see Fig. \ref{dos}. 
The four lowest-lying Zn $4s$ and $4p$ conduction bands are entangled with a somewhat higher-lying dispersive band which has predominantly S $4s$, $4p$ character, see Fig.\ \ref{hfband}. The admixture of S $4s$ and $4p$ states is apparent in the DOS plots in Fig.\ \ref{dos}. 
The fundamental band gap of the system is direct, $\Gamma^{v}_{15} \rightarrow \Gamma^{c}_{1}$, in agreement with experiment \cite{dpos1}. As expected, the HF result, 10.96 eV, strongly overestimates the actual experimental value of 3.83eV \cite{bgzns}. 
The average energy position E$_d$ of the $d$ band complex at the $\Gamma$ point relative to the top of the valence band is --13.3 eV. 
 At the HF level this quantity is overestimated by about 4.0 eV as compared to experimental values of --8.97 eV and --9.03 eV deducted from valence-band x-ray photoemission spectroscopy \cite{dpos1}. 

 As explained above, the orbital basis set employed for our post-HF correlation treatment consists of a set of optimally localized WO's. The Wannier-Boys localization \cite{Zikovich} is carried out separately for the core and valence, and lowest-lying
conduction bands. Because of the finite energy gap between the lower-lying conduction bands and the higher virtual Bloch states, the Wannier-Boys transformation for the former states is straightforward. 

The Wannier-Boys transformation for the four lower-lying Zn $4s$, $4p$ conduction bands yields a set of four $sp^{3}$-like hybrids each oriented along one of the Zn-S segments of a
ZnS$_{4}$ tetrahedron. Each of those hybrids, centered at the Zn site of the ZnS$_{4}$ unit, has significant weight at the nearest-neighbour (NN) S sites and strong antibonding character with the S $3s$ and S $3p$ orbitals. 
The larger lobe of each of the $sp^{3}$-like hybrids is tetrahedrally shaped in the vicinity of the Zn site.
The WO associated with the higher-lying conduction band of predominantly S $4s$, $4p$ character is delocalized over four S sites. 
Our study is restricted here to the correlation-induced corrections for the lower-lying Zn $4s$ and $4p$ states. Electron correlation effects for the S $4s$ and $4p$ bands are not addressed in the following. 

\begin{figure}[htbp]
\begin{center}
\includegraphics[width=8.5cm]{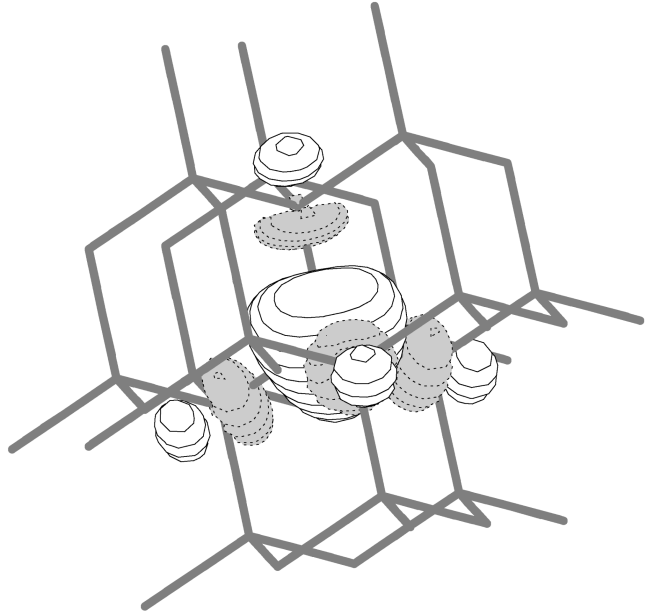}
\caption{Zn $4s$-like conduction-band WO after projection onto a  [Zn$_{13}$S$_{28}$] cluster.}
\label{wo4s}
\end{center}
\end{figure}

\begin{figure}[htbp]
\begin{center}
\includegraphics[width=8.5cm]{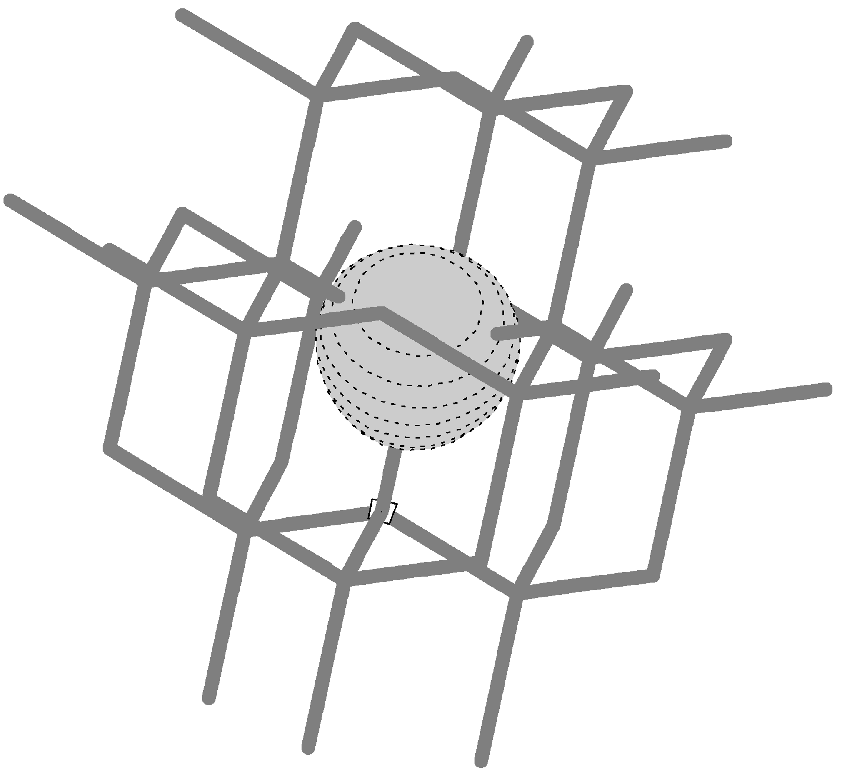}
\includegraphics[width=8.5cm]{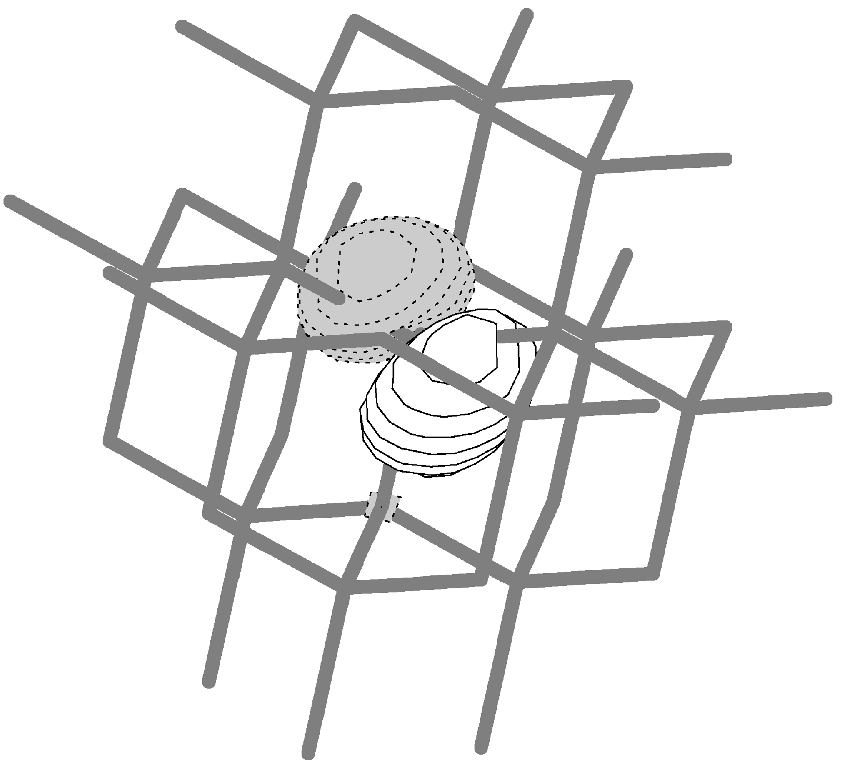}
\caption{S $3p$ and S $3s$ valence-band WO's after projection onto a  [S$_{13}$Zn$_{28}$] cluster.}
\label{wosps}
\end{center}
\end{figure}

In order to obtain a set of Cartesian type $s$ and $p$ functions, a Pipek-Mezey localization procedure \cite{PM} is applied to the projected Zn $sp^{3}$-like hybrids.
\begin{figure}[htbp]
\begin{center}
\includegraphics[width=8.5cm]{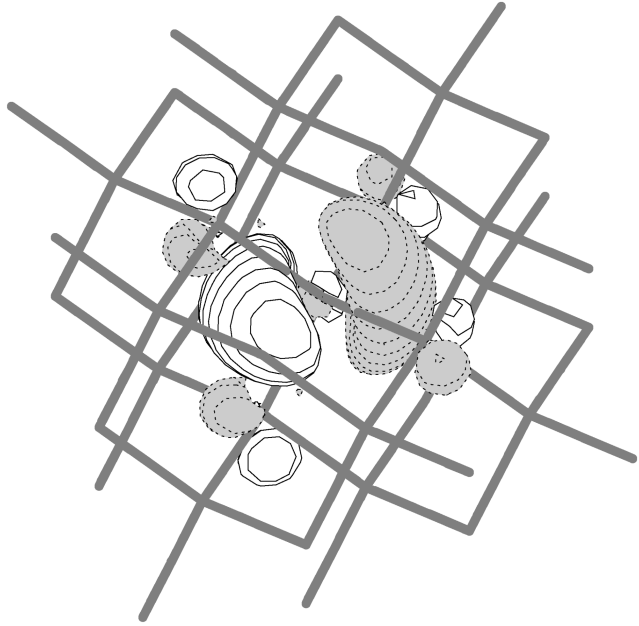}
\caption{Zn $4p$-like conduction-band WO's after projection onto a  [Zn$_{13}$S$_{28}$] cluster.}
\label{wo4p}
\end{center}
\end{figure}
The Pipek-Mezey transformation yields a set of 
three degenerate orbitals, each oriented along one of the $C_2$ axes of the ZnS$_{4}$ tetrahedron and a fourth, lower-energy WO. These projected WO's, derived from the Pipek-Mezey transformation,
are plotted in Figs.\ \ref{wo4p} and \ref{wo4s}.
The degenerate WO's resemble three Cartesian $p$ functions while the fourth WO is 
strongly deformed as compared to a $s$ function, i.e., it is tetrahedrally shaped in the vicinity of the Zn site. All these four Wannier functions have substantial weight at the NN S sites.
Whereas the Zn $4p$-like and Zn $4s$-like WO's are rather extended objects,
the WO's associated with the
sulphur $3s$ and $3p$ valence bands are well localized around the S sites and the small tails at the NN Zn atoms are hardly visible in the plot shown in Fig.\ \ref{wosps}. 
The projected Zn $3s$ and $3p$ WO's obtained through the Pipek-Mezey transformation are slightly deformed Cartesian-like $3s$, $3p_{x}$, $3p_{y}$, and $3p_{z}$ functions. 

Finally, the projected WO's associated with the Zn $3d$ bands are very compact and bear complete resemblance to the $t_2$ ($d_{xy}$, $d_{xz}$, $d_{yz}$) and $e$ ($d_{x^2-y^2}$, $d_{z^2}$) orbitals of a tetrahedrally coordinated Zn ion. 

The norms of the projected valence-band S $3s$, $3p$ and of the semicore Zn $3d$ WO's, centered within the $\mathcal{C}_A$ regions of the embedded clusters, are always larger than 0.99 of the initial HF WO's. The norms of the projected conduction-band WO's associated with the Zn $4s$ and $4p$ states are always larger than 0.95. 

 \subsection{Correlation-induced corrections to the diagonal matrix elements of the local Hamiltonian}
Various  incremental  contributions to the short-range correlation-induced corrections to the diagonal matrix elements of the local Hamiltonian $H^{\textsc{scf}}_{\mathbf{0}n',\mathbf{R}_{I}n}$ are obtained by using different, properly designed clusters. 
Each embedded cluster $\mathcal{C}$ is designed such that a ground-state HF calculation which includes in the SCF optimization only the projected WO's in the $\mathcal{C}_A$ region leads to changes in the total HF ground-state energy of less than few tenths of meV. This demonstrates that the projected WO's centered within the $\mathcal{C}_A$ region are practically identical to the original crystal WO's and that the embedding potentials were properly constructed. 
In other words, the HF description of those finite clusters is indeed equivalent to the HF description of the periodic crystal.
\subsubsection{Short-range relaxation and polarization effects on the ``semicore" Zn $3d$ band states}
\begin{figure}[htbp]
\begin{center}
\includegraphics[width=8.5cm]{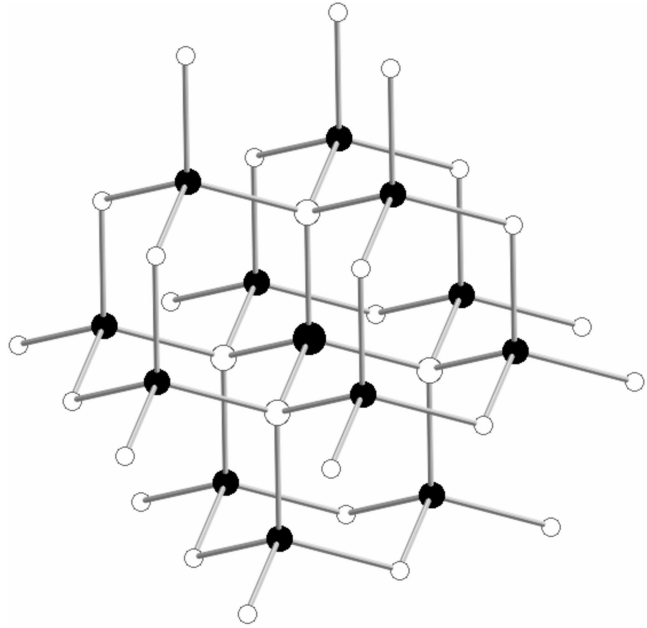}
\caption{Illustration of the [Zn$_{13}$S$_{28}$] cluster employed for the study of the valence-band Zn $3d$ hole states and conduction-band Zn $4s$, $4p$ electron states. The active region of the cluster consists of the [Zn$_{13}$S$_{4}$] fragment. The central Zn site in the active region is depicted as the largest black sphere. The other twelve Zn atoms in the active zone are shown as average-size black spheres. The four S atoms in the active region are drawn as large white spheres, whereas the buffer S atoms are shown as smaller white spheres.}
\label{cluster}
\end{center}
\end{figure}

In order to compute the effect of short-range correlations on the diagonal matrix elements associated with the Zn $3d$ states, we designed a [Zn$_{13}$S$_{28}$] cluster. 
We are here concerned with on-site relaxation plus relaxation and polarization related to the first (four NN S sites) and second (twelve next-nearest-neighbor (NNN) Zn ions) coordination shells around the Zn site which accommodates a $3d$ hole.
The $\mathcal{C}_A$ region, [Zn$_{13}$S$_{4}$], of the [Zn$_{13}$S$_{28}$] cluster contains a central Zn site, depicted as a large black sphere in Fig.\ \ref{cluster}, the four NN S atoms, pictured as large white spheres, and the shell of 12 NNN Zn atoms, drawn as average-size black spheres. The $\mathcal{C}_B$ region includes all S nearest neighbors of the 12 outer Zn sites within the $\mathcal{C}_A$ region. 
The on-site relaxation and polarization effects for the Zn $t_2$ and $e$ ($N\!-\!1\!$) electron states amount to --4.04 eV, see the fourth and fifth columns in Table \ref{diagonalhole}. Those corrections were computed by restricted open-shell HF (ROHF) calculations where the doubly occupied orbitals at the central Zn site hosting the hole are allowed to relax and polarize in response to the missing charge. 
The Zn $3d$ orbital where the hole resides, either of $t_2$ or $e$ symmetry, is kept frozen. 
\begin {table}[h]
 \centering 
 \caption{Correlation-induced corrections (in eV) to the diagonal Hamiltonian matrix elements for the valence-band S $3p$, $3p$ and Zn $3d$ hole states. Negative corrections indicate an upward shift of the valence bands.}
\begin{tabular}{p{5.0cm}p{1.8cm}p{1.8cm}p{1.8cm}p{1.8cm}}
  \hline\hline
 &$\Delta$ H$_{mm}$($\textbf{0}$)&& &\\ \hline 
&S $3s$&S $3p$&Zn $3d_e$ & Zn $3d_{t_{2}}$\\ \hline
On-site orbital relaxation&--1.303&--0.852&--4.031&--4.036\\
NN orbital relaxation &--0.125&--0.116&--2.908&--2.908\\
NNN orbital relaxation & --1.261& --1.181&--0.086& --0.086\\
Hole orbital relaxation &--0.096&--0.155&--0.249& --0.251\\
Loss of GS correlation &1.760&0.822&1.455&1.599 \\
Long-range polarization &--1.537&--1.537&--1.537&--1.537\\
Total correlation correction$^{1}$ &--2.562&\textbf{-3.019}&\textbf{-7.356}&\textbf{-7.219} \\ \hline
 \hline \hline
 \end{tabular} 
\label{diagonalhole}

$^{1}$ differential correlation effects other than loss of ground-state correlations are not included
\end {table}
A modest contribution of about --0.24 eV to the overall on-site correlation effects arises from the relaxation and polarization of the valence Zn $3s$, $3p$ WO's. As anticipated, the largest effect comes from the relaxation of the doubly occupied Zn $3d$ WO's, that is --3.79 eV. 
The $3d$ hole states in first row transition metal (TM) compounds are characterized by large atomic relaxation effects due to the core-like spatial distribution of the $3d$ orbitals. 

Orbital relaxation and polarization at the NN S sites are also substantial, --2.91 eV, see Table \ref{diagonalhole}. 
Relaxation and polarization within the first zinc coordination shell of the central Zn ion result into 
a small additional correction of about --0.09 eV. Finally, the relaxation of the singly occupied Zn $t_{2}$ or $e$ WO where the hole itself resides brings an extra correction of --0.25 eV. 

As illustrated in Table \ref{diagonalhole}, the short-range relaxation effects are nearly the same for the Zn $t_2$ and $e$ ($N\!-\!1\!$) states. This is due to the very compact nature of the Zn $3d$ orbitals. This compact character of the Zn $3d$ orbitals is also the reason for a crystal field splitting that amounts to only 0.01 eV. Correlation effects beyond relaxation and polarization actually enhance the crystal field splitting to about 0.1 eV, see Section IV. B. 3.
The small value of the Zn $t_{2}$-$e$ splitting is consistent with the experimental findings, i.e., no splitting of the peak corresponding to the Zn $3d$ semicore states is observed in the x-ray photoelectron spectrum of ZnS \cite{dpos1, Langer}. 

In the calculations above, we constructed the relaxed ROHF wave functions $|\tilde{\Phi}^{N-1}_{\mathbf{R}_{I}m\sigma}\rangle$ as first approximations to the correlated wave functions $|\Psi^{N-1}_{\mathbf{R}_{I}m\sigma}\rangle$ starting from the Koopmans electron-removal states $|\Phi^{N-1}_{\mathbf{R}_{I}m\sigma}\rangle$.
The influence of scalar relativistic effects and spin-orbit coupling on the Zn $3d$ shell is not explicitly considered. 
The spin-orbit splitting within the $3d$ shell of the Zn$^{3+}$ ion is 0.34 eV, much larger than the crystal-field splittings. The effect of those spin-orbit couplings, however, is much smaller than the relaxation and polarization effects which determine the actual binding energies of the Zn $3d$ semicore states.  
Relaxation and polarization effects beyond the first zinc coordination shell are evaluated via a dielectric continuum approximation, see below. 

The small radial extent of the Zn $3d$ orbitals and the large separation between the Zn $3d$ and S $3p$ energy levels leads actually to small mixing between the Zn $3d$ and S $3p$ functions. Ligand $p$ to metal $d$ charge transfer (CT) effects are therefore expected to be very small in ZnS. 
Trial multiconfiguration, complete active space (CAS) SCF calculations \cite{Helgaker} for the 
$3d$ ($N\!-\!1\!$) electron states were also performed, where the central Zn $3d$ orbitals and the twelve S $3p$ orbitals centered within the $\mathcal{C}_A$ region were all included in the active orbital space. These calculations show indeed that the weight of S $3p$ to Zn $3d$ CT configurations in the CASSCF wave functions is not larger than 1 \%. The energy separation between the highest-lying S $3p$ ($N\!-\!1\!$) state and the lowest-lying Zn $3d$ ($N\!-\!1\!$) state is also found to be substantial, 4.5 eV. 
The optimized active orbitals turn out to constitute a set of optimally localized atomic-like active orbitals with small mixing between the S $3p$ and Zn $3d$ basis functions, which indicates weak covalency in the Zn $3d$ ($N\!-\!1\!$) states. Photoemission studies by Langer and Vesely \cite{Langer} also indicate relatively small covalency of the Zn-S bond.   
A ROHF treatment of the type described above provides then a good description for the Zn $3d$ and S $3p$ ($N\!-\!1\!$) states. 
\subsubsection{Short-range relaxation and polarization effects on the valence S $3s$, $3p$ band states}

Relaxation and polarization effects in the nearest neighbourhood of a sulphur $3s$ or $3p$ hole are also assessed by means of ROHF calculations on a [S$_{13}$Zn$_{28}$] cluster. 
The cluster [S$_{13}$Zn$_{28}$] is designed in perfect analogy with the [Zn$_{13}$S$_{28}$] cluster, with an active region [SZn$_{4}$] containing a central S ion, where the hole resides, and its four NN Zn ions. 
The wave functions $|\tilde{\Phi}^{N-1}_{\mathbf{R}_{I}m\sigma}\rangle$ for the S $3s$, $3p$ ($N\!-\!1\!$) states are constructed by allowing for all valence and core orbitals within the $\mathcal{C}_A$ region to relax and polarize in  response to the extra charge attached to the S $3p$ or $3s$ WO.

The corrections to the frozen orbital (Koopmans') ionization energies due to on-site orbital relaxation turn out to be substantial, 0.85--1.30 eV, for the S $3p$ and $3s$ hole states, see second and third columns in Table \ref{diagonalhole}. 
As expected, the net effect of the orbital relaxation is larger for the S $3s$ hole states due to the fact that all six electrons in the polarizable S $3p$ shell are readjusting.  

The orbital relaxation and polarization effects at the NN Zn sites give rise to an additional moderate correction of about --0.12 eV,  
see fourth row in Table \ref{diagonalhole}.
Such a moderate effect reflects the strongly localized, core-like character of the Zn WO's.

\begin{figure}[htbp]
\begin{center}
\includegraphics[width=8.5cm]{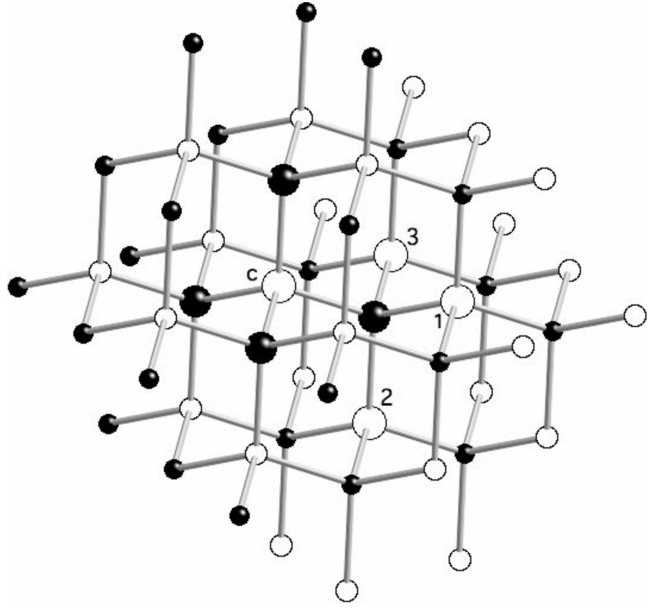}
\caption{Illustration of the [S$_{28}$Zn$_{28}$] cluster employed for the study of the relaxation and polarization effects within the first sulphur coordination shell around a valence-band S $3p$, $3s$ hole. The active region of the cluster is the [S$^{c}$Zn$_{4}$S$_{3}$] fragment, 
where with the superscript is marked the central S$^{c}$ ion. The other three active S ions in unit cells with $\mathbf{R}_{1}$, $\mathbf{R}_{2}$, and $\mathbf{R}_{3}$, and the S$^{c}$ ion are depicted as large white spheres. The four active Zn ions are drawn as large black spheres. The S ions in the buffer region are shown as smaller white spheres, whereas the buffer Zn atoms are drawn as smaller black spheres.}
\label{clusterS28Zn28}
\end{center}
\end{figure}

Relaxation and polarization effects within the first sulphur coordination shell of the central S site are also explicitly considered. This coordination shell contains 12 NNN S sites. The distances between each of the twelve NNN sulphurs and the central S site are 3.829 \AA. Wave-function-based calculations for a cluster containing all twelve NNN S sites are computationally unfeasible even with the smaller basis set described below. To evaluate the overall relaxation effect associated with the 12 NNN S ions, we designed a [S$_{28}$Zn$_{28}$] cluster containing only three of the twelve NNN S sites.
The $\mathcal{C}_A$ region [S$^{c}$Zn$_{4}$S$_{3}$] of the [S$_{28}$Zn$_{28}$] cluster consists of the 
$\mathcal{C}_A$ region [S$^{c}$Zn$_{4}$] of the [S$_{13}$Zn$_{28}$] cluster and those three sulphur sites situated in unit cells with lattice vectors $\mathbf{R}_{1}=(1,1,0)\frac{a}{2}$, $\mathbf{R}_{2}=(1, 0, -1)\frac{a}{2}$, and $\mathbf{R}_{3}=(0, 1,-1)\frac{a}{2}$, see Fig.\ \ref{clusterS28Zn28}.
They are positioned at equal distances with respect to each other and with respect to the central S$^{c}$ site, i.e., 3.829 \AA.  
The associated buffer region $\mathcal{C}_B$ contains the remaining 9 NNN sulphur ions, 
all zinc ions in the first coordination shell of the 12 NNN S sites, and all sulphurs in the second coordination shell of the selected three active NNN S sites. 
The 86-311G basis set described in Sec. III is employed for S where the polarization $d$ basis function is, however, omitted. The basis set utilized for Zn is derived from the 86-4111d41G GTO basis set of Jaffe \textit{et al.} \cite{Jaffe} by optimizing the exponents of the outermost $4sp$, $5sp$ and $6sp$ shells. 
This optimization does not alter their values as compared to the 86-4111d311G GTO basis described in Sect. III. The exponents and contraction coefficients of the $3d$ and $4d$ shells in the basis of Jaffe \textit{et al.} appear to be optimal and undergo no variations in the basis set optimization. 

Orbital relaxation and polarization at only one of the three NNN S ions bring extra corrections of about --0.11 eV for both the S $3s$ and $3p$ hole states. This correction may be viewed as the relaxation-energy one-body increment associated with a single NNN S site in a many-body expansion for the overall relaxation contribution of the twelve NNN S ions \cite{Stoll1}. 

Basis set effects were found to be negligibly small, less than 0.005 eV, from additional calculations for the on-site and NN orbital relaxation and polarization on the [S$_{13}$Zn$_{28}$] cluster and using either the 86-4111d41G or 86-4111d311G GTO basis sets for Zn and the 86-311G* basis set for S. 
The impact of the polarization $d$ function for sulphur on the relaxation and polarization effects at the NNN S sites is also explicitly investigated by computing the one-body increment to the overall relaxation effect of the twelve NNN S ions with a [S$_{20}$Zn$_{28}$] cluster, the 86-4111d41G basis set for zinc and either the 86-311G* or 86-311G basis sets for sulphur. 
The cluster [S$_{20}$Zn$_{28}$] contains only one of the three NNN S sites included in the [S$_{28}$Zn$_{28}$] cluster. The active region [S$^{c}$Zn$_{4}$S$_{1}$] consists of the $\mathcal{C}_A$ region [S$^{c}$Zn$_{4}$] of the [S$_{13}$Zn$_{28}$] cluster and the sulphur ion situated in unit cell with lattice vector 
$\mathbf{R}_{1}=(1, 1, 0)\frac{a}{2}$, see Fig.\ \ref{clusterS28Zn28}. The buffer region contains the other eleven NNN S as well as all Zn ions in the first coordination shell of the twelve sulphurs. The sulphur ions in the second coordination shell of the NNN active site S$_1$  are also included in the $\mathcal{C}_B$ region.
By utilizing either the 86-311G* or 86-311G basis sets for sulphur, the orbital relaxation and polarization at the NNN active site S$_1$ give rise to a one-body increment of --0.12  or --0.11 eV, respectively for both the S $3s$ and $3p$ hole states. Hence, the basis set effect is found to be 
about 0.01 eV. Such basis set effects are incorporated in all numbers discussed below. 

Relaxation and polarization effects at a second sulphur site in the [S$_{28}$Zn$_{28}$] cluster give rise to an extra correction of --0.097 eV for the S $3p$ ($N\!-\!1\!$) states and --0.100 eV for the S $3s$ ($N\!-\!1\!$) states. 
Finally, when the orbitals at the third NNN S ion in the $\mathcal{C}_A$ region [S$^{c}$Zn$_{4}$S$_{3}$] are  allowed to relax and polarize, an additional small contribution arises that amounts to --0.076 eV for the S $3p$ ($N\!-\!1\!$) and --0.091 eV for the S $3s$ ($N\!-\!1\!$) states. 
Summing up the relaxation and polarization contributions of these three NNN S sites situated at equal distances of 3.829 \AA $\,$ with respect to each other, we obtain an overall effect of --0.295 eV for the S $3p$ and --0.315 eV for the S $3s$ ($N\!-\!1\!$) states. The overall orbital relaxation and polarization effect at the twelve NNN S sites is then calculated by multiplying by four the overall contribution of the three NNN S sites discussed above. The total orbital relaxation contributions are listed in Table \ref{diagonalhole}, fifth row, second and third columns. 
A comparison of those contributions for the S $3s$ and $3p$ ($N\!-\!1\!$) states reveals their small dependence of the angular momentum of the orbital hosting the hole. The interaction between the extra charge at the central S and the induced dipole at the NNN S sites is thus mainly a monopole-dipole type interaction.

It is interesting to consider the relaxation and polarization effects at two NNN S sites which lie with respect to each other at a distance larger than 3.829 \AA. In the crystal structure of $c$-ZnS, this distance is either 5.415, 6.632 or 7.658 \AA $\,$. For this purpose, we constructed the [S$_{26}$Zn$_{28}$] and [S$_{27}$Zn$_{28}$] clusters, respectively. 
The active regions [S$^{c}$Zn$_{4}$S$_{2}$] of the two clusters contain the [SZn$_{4}$] fragment and the NNN S ion located in the unit cell with $\mathbf{R}_{1}=(1, 1, 0)\frac{a}{2}$. 
The second NNN S ions included in the $\mathcal{C}_A$ regions of the [S$_{26}$Zn$_{28}$] and [S$_{27}$Zn$_{28}$]  clusters are situated in unit cells with $\mathbf{R}_{4}=(-1,1,0)\frac{a}{2}$ and $\mathbf{R}_{5}=(-1,0,1)\frac{a}{2}$, respectively. The distance between the two NNN S ions in each of the two clusters is 5.415 \AA $\,$ and 6.632 \AA $\,$, respectively.
The buffer regions of the two clusters are built in perfect analogy with that of the [S$_{28}$Zn$_{28}$] cluster. All calculations were carried out with the basis sets used for the  [S$_{28}$Zn$_{28}$] cluster. 

Calculations for the one-body increment discussed above exploiting either the [S$_{26}$Zn$_{28}$], [S$_{27}$Zn$_{28}$] or [S$_{28}$Zn$_{28}$] clusters show that this quantity is converged with respect to the cluster size, i.e., a negligible increase by less than 0.003 eV is observed from the clusters [S$_{26}$Zn$_{28}$] and [S$_{27}$Zn$_{28}$] to the cluster [S$_{28}$Zn$_{28}$].

Orbital relaxation and polarization effects at a second NNN S site which is positioned at the distance d$_{SS}$=5.415 \AA \ with respect to a S site selected arbitrary to be the first S site, as in the cluster [S$_{26}$Zn$_{28}$], bring a correction of --0.105 eV for the S $3p$ and  --0.107 eV for the S $3s$ ($N\!-\!1\!$) states. 
When the second sulphur ion is situated at d$_{SS}$=6.632 \AA, as in the cluster [S$_{27}$Zn$_{28}$], the extra contribution to the overall relaxation and polarization amounts to --0.106 eV for the S $3p$ and --0.109 eV for the S $3s$ ($N\!-\!1\!$) states. 
These findings demonstrate that the orbital relaxation contributions associated with two distinct NNN S sites are approximately additive at distances between them larger than 3.829 \AA. The non-additive corrections to the overall relaxation of the joint orbital system of the two NNN  S ions are small, about --0.01 eV.
We have not explicitly considered the orbital relaxation effect at a second NNN S site situated at d$_{SS}$=7.658 \AA \ with respect to the first S site. The orbital relaxation contributions associated with two sulphurs at this distance are expected to be nearly additive, $\approx$ --0.24 eV.

Overall contributions due to charge relaxation at three NNN S sites with distances between any two of them larger than 3.829 \AA $\,$ are not explicitly computed. 
Following a similar line of reasoning as for the charge relaxation at two NNN S sites, those contributions are expected to be about --0.3 eV.   

Finally, there is an additional contribution due to the relaxation of the orbital hosting the hole of --0.1 eV, see Table \ref{diagonalhole}.
Summing up the on-site and short-range relaxation and polarization contributions, we obtain an overall correlation-induced shift of the S $3p$ valence bands of --2.30 eV. 
\subsubsection{Short-range relaxation and polarization effects on the low-lying conduction Zn $4s$, $4p$ band states}
In the following, we focus on the relaxation and short-range polarization effects on the diagonal matrix elements for the low-energy Zn $4s$, $4p$ ($N\!+\!1\!$) states. These electron-addition states are associated with configurations with an extra electron into the Zn $4s$-like or Zn $4p$-like conduction-band WO's.
To investigate those configurations, we employed a [Zn$_{13}$S$_{28}$] cluster with an active region [Zn$_{13}$S$_{4}$] as in the study of the Zn $3d$ semicore states. The basis sets exploited are the 86-4111d311G basis set for Zn and the 86-311G* basis set for S.
In perfect analogy to the S $3s$, $3p$ ($N\!-\!1\!$) valence-band states, ROHF calculations were performed for the lowest-energy Zn $4s$, $4p$ ($N\!+\!1\!$) states keeping the singly-occupied Zn $4s$ or Zn $4p$ orbitals frozen \cite{Pahl}. The on-site orbital relaxation and polarization in response to the extra electron hosted by either the Zn $4s$ or Zn $4p$ WO's amount to about --0.05 eV, see Table \ref{diagonalelec}, first row. 
\begin {table}
 \centering 
 \caption{Correlation-induced corrections (in eV) to the diagonal Hamiltonian matrix elements for the conduction-band Zn $4s$ and Zn $4p$ added-electron states. Negative corrections indicate a downward shift of the conduction bands.}
\begin{tabular}{p{5.0cm}p{1.9cm}p{1.9cm}}
  \hline\hline
 &$\Delta$ H$_{nn}$($\textbf{0}$)& \\ \hline 
&Zn $4s$-like&Zn $4p$-like\\ \hline
On-site orbital relaxation&--0.068&--0.043\\
NN orbital relaxation&--0.908&--0.785\\
NNN orbital relaxation &--0.073 &--0.073 \\
Loss of GS correlation &+0.130& +0.077\\
Long-range polarization &-1.537&-1.537\\
Total correlation correction &\textbf{-2.456}& \textbf{-2.361}\\ 
 \hline \hline
 \end{tabular} 
\label{diagonalelec}
\end {table}
This small correction to the diagonal matrix elements for the Zn $4s$, $4p$ ($N\!+\!1\!$) states 
is consistent with the core-like character of the valence Zn $3s$, $3p$, and $3d$ orbitals, and the small polarizability of the Zn$^{2+}$ ion. 
Relaxation and polarization effects at the NN sulphur ions contribute to a downward shift of about 0.8--0.9 eV for the Zn $4s$ and $4p$ conduction bands, see Table \ref{diagonalelec}. Orbital relaxation at the NNN Zn sites brings an additional small downward shift of about 0.07 eV of the center of gravity of the bands.

In contrast to the S $3s$, $3p$ and Zn $3d$ ($N\!-\!1\!$) states, where the WO's hosting the hole 
did not delocalize during the SCF optimization, 
for the Zn $4s$ and $4p$ ($N\!+\!1\!$) states, the localization of the extra electron in the 
Zn $4s$-like or Zn $4p$-like orbitals could not be maintained even by freezing higher-energy $s$ and $p$ virtual orbitals at the nearest S and Zn sites. This particular effect is therefore neglected for the conduction-band states.  

\subsubsection{Long-range polarization effects}
An important contribution to the shift of the center of gravity of the energy bands and hence to the reduction of the HF gap comes from long-range polarization effects. To obtain an estimate of the long-range polarization effects, we compute the classical polarization energy of a dielectric medium outside a sphere of radius $R$ beyond which the dielectric response of the crystal reaches its asymptotic value $\epsilon_0$ \cite{Fuldebook}:
 \begin{eqnarray}
\triangle E (R)=-\frac{\epsilon_0-1}{2\epsilon_{0}}\frac{e^{2}}{R}.
\nonumber
 \end{eqnarray}
The dielectric continuum approximation is a well suited approximation for the evaluation of long-range polarization effects beyond a given $R$, since the interaction of the extra charge with the surroundings beyond the radius $R$ has predominantly electrostatic character, see, e.g., Refs. ~\onlinecite{Hozoi, AlexJCP}. We make use of the static dielectric constant $\epsilon_0$=8.34 deduced by Hattori \textit{et al.} \cite{Hattori} from optical data. Corrections of few eV were found to the diagonal matrix elements for the valence- and conduction-band states. Those corrections are included in Tables \ref{diagonalhole} and \ref{diagonalelec} as well. 
As discussed above, relaxation and short-range polarization effects within the sphere of radius $R$
 were computed by \textit{ab initio} methods. 

The cuttoff radii $R$ for the Zn $3d$ ($N\!-\!1\!$) and $4s$, $4p$ ($N\!+\!1\!$) excited states are each obtained as the average of the radii of the first and third sulphur coordination shells around the central Zn site. Likewise, the cuttoff radii for the S $3s$, $3p$ ($N\!-\!1\!$) states are obtained as the average of the radii of the first and third zinc coordination shells around the central S site. Because of the equivalent structure of the coordination shells around the Zn and S sites, the corresponding values for the cuttoff radii $R$ for the valence-band hole states and for the conduction-band electron-addition states are the same, 4.123 \AA $\,$. Thus, at this level of approximation the long-range polarization correction $\Delta H^{lr}_{mm (nn)}(\mathbf{0})$ for each of the two types of states is  --1.54 eV.

The long-range polarization contributions caused by an extra charge in the conduction or valence bands of an insulator may be taken beyond a certain radius $R$ to be symmetric, since the polarizing charges have the same absolute value. 
This does not hold for the short-range relaxation and polarization contributions (see Tables \ref{diagonalhole} and \ref{diagonalelec}) because at short distances the associated charge distribution and polarization can not be approximated by that of a dielectric continuum. 
\subsubsection{Differential correlation effects: loss of ground-state correlations}

In addition to charge relaxation and polarization, a relevant contribution to the correlation-induced corrections to the diagonal matrix elements also arises from differential correlation effects. An important fraction corresponds here to the loss of ground-state correlations. 
To study such effects, we construct the correlated wave functions $|\Psi^{N-1}_{\mathbf{R}_{I}m\sigma}\rangle$ and $|\Psi^{N+1}_{\mathbf{R}_{I}n \sigma}\rangle$ via SDCI calculations starting from the reference wave functions $|\breve{\Phi}^{N-1}_{\mathbf{R}_{I}m\sigma}\rangle$ and $|\tilde{\Phi}^{N+1}_{\mathbf{R}_{I}n\sigma}\rangle$, respectively. 
The clusters [S$_{13}$Zn$_{28}$] and [Zn$_{13}$S$_{28}$] are employed for the study of correlation effects beyond charge relaxation and polarization for the S $3s$, $3p$ ($N\!-\!1\!$) and Zn $3d$ ($N\!-\!1\!$) or Zn $4s$, $4p$ ($N\!+\!1\!$) states, respectively. The active region of the [S$_{13}$Zn$_{28}$] cluster is the [SZn$_{4}$] fragment, while that of the [Zn$_{13}$S$_{28}$] cluster consists of a [Zn$^c$S$_{4}$Zn$_{12}$] fragment. 
The basis sets utilized in the SDCI calculations are the 86-311G* basis set for S and the 86-4111d311G basis for the Zn ions.

To calculate the loss of ground-state correlations, we utilized either the total SDCI correlation energies or relevant fractions thereof for the $N$-electron ground state and for the Zn $3d$, S $3s$, $3p$ ($N\!-\!1\!$) states or Zn $4s$, $4p$ ($N\!+\!1\!$) states. For the $N$-electron GS and Zn $3d$ ($N\!-\!1\!$), and Zn $4s$, $4p$ ($N\!+\!1\!$) states, the SDCI wave functions are constructed by correlating the $3s$, $3p$, and $3d$ orbitals of the Zn$^c$ ion in the $\mathcal{C}_A$ region [Zn$^c$S$_{4}$Zn$_{12}$] plus the $3s$ and $3p$ orbitals of the four NN S ions. For the S $3s$ and $3p$ ($N\!-\!1\!$) and $N$-electron states as obtained from the [S$_{13}$Zn$_{28}$] cluster, the S $3s$, $3p$ and Zn $3d$ orbitals in the active region [SZn$_4$] are explicitly correlated. 
The results for the loss of ground-state correlations are reported in Tables \ref{diagonalhole} and \ref{diagonalelec}. The corrections to the diagonal matrix elements of the effective Hamiltonian due to that effect shift the Zn $3d$ bands downward by about 1.5-1.6 eV. A similar correction of about 1.0 eV  was also found for the S $3p$ band states. The Zn $4s$ and $4p$ electron-addition states are shifted upwards by about 0.1 eV due to the loss of ground-state correlations.
In order to understand the net results reported here for the loss of ground-state correlations, some technical explanations are required
without which the paper is not self-contained. They are found in Appendix
B and should enable the reader to repeat calculations of the type
presented here.

\subsubsection{Overall correlation-induced corrections to the band energies}
Summing up all relevant correlation-induced corrections to the diagonal matrix elements for the valence-band states, the average energy position of the Zn $3d$  semicore states relative to the top of the valence S $3p$ bands at the $\Gamma$ point is found to be --8.91--9.01 eV, in very good agreement with experimental values of --8.97 and --9.03 eV extracted from valence-band x-ray photoemission \cite{dpos1}.

The overall effect of the correlation-induced corrections considered in the current study to the fundamental gap is a reduction of the HF value from 10.96 eV to 5.49 eV, which is still 1.69 eV larger than the experimental estimate of 3.83 eV \cite{bgzns}.  We do not expect that more sophisticated basis sets would significantly affect 
 those quantities.
\subsection{Off-diagonal matrix elements of the local effective Hamiltonian}
In the following paragraphs, we consider the correlation-induced corrections to the band widths. To obtain these corrections, the off-diagonal matrix elements of the effective Hamiltonian in Eq. (\ref{epsiloncorr2})
are explicitly computed. We apply thereby the computational approach adopted in Refs. ~\onlinecite{Hozoi, AlexJCP}. 

The effective Hamiltonian matrix elements between orthogonal frozen-orbital ROHF wave functions $|\Phi^{N-1}_{\mathbf{R}_{I}m\sigma}\rangle$ and $|\Phi^{N-1}_{\mathbf{0}m'\sigma}\rangle$ having the hole at distinct sites are explicitly computed. 
These off-diagonal matrix elements $H^{\textsc{scf}}_{\mathbf{R}_{I}, mm'}$=$\langle \Phi^{N-1}_{\mathbf{R}_{I}m\sigma}|H|\Phi^{N-1}_{\mathbf{0}m'\sigma}\rangle$ are directly related to the hopping terms of an orthogonal tight-binding representation. 
The same considerations hold for the wave functions  $|\Phi^{N+1}_{\mathbf{R}_{I}n\sigma}\rangle$ and matrix elements $H^{\textsc{scf}}_{\mathbf{R}_{I}, nn'}$.

Relaxation and polarization effects in the nearby surroundings of the added hole or electron are explicitly incorporated by means of separate SCF optimizations for each of the ($N\!-\!1\!$) and ($N\!+\!1\!$) electron states. 
The correlated wave functions $|\tilde{\Phi}^{N-1}_{\mathbf{R}_{I}m\sigma}\rangle$, $|\tilde{\Phi}^{N-1}_{\mathbf{0}m'\sigma}\rangle$ and $|\tilde{\Phi}^{N+1}_{\mathbf{R}_{I}n\sigma}\rangle$, $|\tilde{\Phi}^{N+1}_{\mathbf{0}n'\sigma}\rangle$ are thereby constructed.
These wave functions are each expressed in terms of its own orbital set and are thus mutually non-orthogonal. 
The calculation of the Hamiltonian and overlap matrix elements between such nonorthogonal wave functions is performed utilizing an approach that has been implemented in {\sc molpro} by Mitrushchenkov and Werner and rests on non-unitary transformations of the initial set of nonorthogonal orbitals to biorthogonal sets \cite{Mitrushchenkov}.

To construct the correlated energy bands $\epsilon_{\mathbf{k}\mu\sigma}$ and $\epsilon_{\mathbf{k}\nu\sigma}$, one can opt for diagonalizing a $\mathbf{k}$-dependent matrix like that in Eq.\ (\ref{epsiloncorr2}), which contains the Hamiltonian and overlap matrix elements between the non-orthogonal, correlated wave functions $|\tilde{\Phi}_{\mathbf{R}_{I}}^{N\pm1}\rangle$ or $|\Psi_{\mathbf{R}_{I}}^{N\pm1}\rangle$. 
An alternative route is based on deriving from the initial inter-site Hamiltonian and overlap matrix elements a set of effective hopping integrals for an orthogonal tight-binding-like formulation. In the latter approach, the effective hopping terms can be directly compared with the HF hoppings which offers 
a better insight into the effect of correlations on the electronic band structure.

For mutually non-orthogonal electron-removal states with different binding energies, i.e., $H_{\mathbf{0}, m'm'} \neq H_{\mathbf{R}_{I}, mm}$ and $S_{\mathbf{R}_{I}, mm'}\neq 0$, the hopping terms read \cite{AlexJCP}
 \begin{eqnarray}
 \nonumber
 t_{mm'} (\mathbf{R}_{I})&=&\frac{1}{1-S^{2}_{\mathbf{R}_{I}, mm'}}\Bigg\{H_{\mathbf{R}_{I}, mm'}-
 \nonumber   \\
 &-&S_{\mathbf{R}_{I}, mm'}\bigg\{\frac{H_{\mathbf{0}, m'm'}+H_{\mathbf{R}_{I}, mm}}{2}\bigg\}\Bigg\}
 \nonumber \\
 &=& \frac{1}{2}\Bigg\{\triangle E^{2}-\frac{(H_{\mathbf{R}_{I}, mm}-H_{\mathbf{0}, m'm'})^{2}}{1-S^{2}_{\mathbf{R}_{I}, mm'}}\Bigg\}^{\frac{1}{2}}.
 \nonumber
\end{eqnarray}
An analogous expression holds for $t_{nn'} (\mathbf{R}_{I})$. $\triangle E$ denotes here the energy separation between the two eigenstates of the 2 x 2 non-orthogonal CI (NOCI) secular problem \cite{Broer}. 
Equivalently, the effective hopping term can also be written as:
\begin{eqnarray}
 \nonumber
 t_{mm'} (\mathbf{R}_{I})&=&H_{\mathbf{R}_{I}, mm'} 
 \nonumber \\
 &=&\frac{1}{2}\Bigg\{\triangle E^{2}-(H_{\mathbf{R}_{I}, mm}-H_{\mathbf{0}, m'm'})^{2}\Bigg\}^{\frac{1}{2}}
 \nonumber
\end{eqnarray}
when $\langle\tilde{\Phi}_{\mathbf{0}m'\sigma}^{N-1}|\tilde{\Phi}_{\mathbf{R}_{I}m\sigma}^{N-1}\rangle$=0.

\begin{figure}[htbp]
\begin{center}
\includegraphics[width=8.5cm]{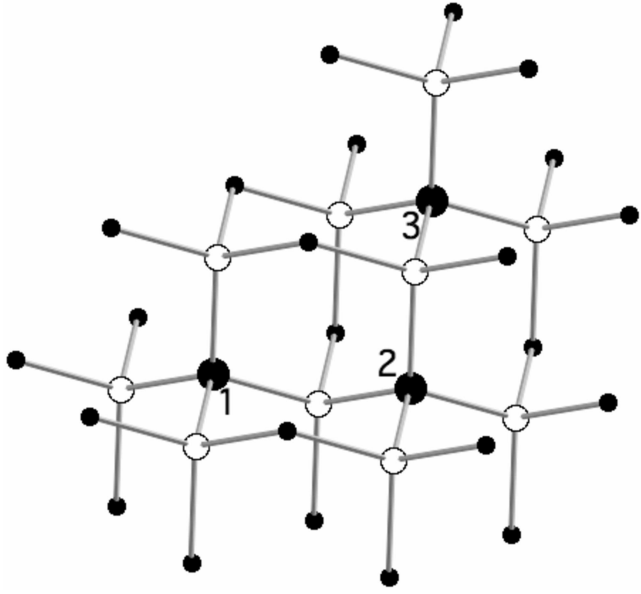}
\caption{A sketch of the [S$_{26}$Zn$_{10}$] embedded cluster employed for the computation of the NNN effective hoppings $t_{mm'}$ associated with the valence-band S $3s$, $3p$ hole states. The $\mathcal{C}_A$ region of the cluster consists of the [S$_{3}$Zn$_{10}$] fragment. The three S sites in the $\mathcal{C}_A$ region are depicted as large black spheres and labeled 1, 2, 3. S ions in the buffer zone $\mathcal{C}_B$ are shown as small black spheres. The ten Zn ions in the active region are drawn as large white spheres.}  
\label{clusterhop}
\end{center}
\end{figure}

For computing NN and NNN effective hopping terms associated with the S $3s$, $3p$ valence bands, two different clusters were designed, [S$_{20}$Zn$_{7}$] and [S$_{26}$Zn$_{10}$]. 
The active regions $\mathcal{C}_A$ of these clusters are the [S$_{2}$Zn$_{7}$] and [S$_{3}$Zn$_{10}$] fragments, respectively. The [S$_{2}$Zn$_{7}$] fragment consists of two S sites, denoted 1 and 2 in Fig.\ \ref{clusterhop}, and their NN zinc ions. 
Likewise, the [S$_{3}$Zn$_{10}$] active region contains the S sites 1 and 2 plus an additional S ion denoted as 3 in Fig.\ \ref{clusterhop}. All NN Zn ions of these three S sites are included in the $\mathcal{C}_A$ subunit as well.
The buffer regions of the two clusters consist of all S ions within the first coordination shell of the active zinc sites. 
The [S$_{26}$Zn$_{10}$] cluster is sketched in Fig.\ \ref{clusterhop}.

The basis sets used in the calculations for the [S$_{20}$Zn$_{7}$] and [S$_{26}$Zn$_{10}$] clusters are 86-311G* for the S species and 86-4111d311G for Zn. 
An overview of the valence-band NN and NNN effective hopping terms is provided in Table \ref{Hop1}. 
\begin {table}
 \centering 
  \caption{Nearest-neighbor (NN), $\mathbf{R}_{I}=(0, -1, 1)\frac{a}{2}$, and next-nearest-neighbor (NNN), $\mathbf{R}_{I}=(0, -1, 0) a$, effective hoppings for the valence-band S $3s$, $3p$ states (in eV).  Frozen-orbital CI results (FOCI ) are listed in the second column. NOCI matrix elements in terms of relaxed orbitals are provided in the third column (ROH-NOCI).  Matrix elements smaller than 0.004 eV are not included in the table.}
  \begin{tabular}{p{2.8cm}p{2.3cm}p{2.3cm}}
  \hline\hline
Active WO's &\multicolumn{2}{c}{NN t$_{mm'}$($\mathbf{R}_{I}$)} \\ 
&FOCI&ROH-NOCI\\ \hline
$3s$-$3s$ &0.080 &0.096 \\
$3p$$_x$-$3p$$_x$ &0.060 &0.065 \\
$3p$$_{y (z)}$-$3p$$_{y (z)}$&0.288&  0.315\\
$3p$$_{y(z)} $-$3p$$_{z (y)}$& 0.368&0.405 \\
$3p$$_{x (y)}$-$3p$$_{y (x)}$&0.079& 0.088\\
$3p$$_{x (z)}$-$3p$$_{z (x)}$&0.079&0.088 \\
\hline
  \hline
Active WO's &\multicolumn{2}{c}{NNN t$_{mm'}$($\mathbf{R}_{I}$)} \\ 
& FOCI&RO-NOCI\\ \hline
$3p$$_{y}$-$3p$$_{y}$&0.018& 0.020 \\
$3p$$_{z}$-$3p$$_{z}$&0.006& 0.008 \\
$3p$$_{y (z)}$-$3p$$_{z (y)}$&0.005 & 0.004\\
\hline\hline
\end{tabular} 
\label{Hop1}
\end {table}
In the first column of Table \ref{Hop1} the Wannier functions where the hole resides are given. The second column summarizes results from frozen-orbital CI calculations in terms of the HF WO's (FOCI). 
In the third column, we allow for the relaxation and polarization of the doubly occupied orbitals of the S and Zn ions within the $\mathcal{C}_A$ regions. 
Additionally, the hole orbital is also allowed to relax within the adjusted environment (ROH-NOCI), as described in Section II.

The orbital relaxations are carried out separately for each electron-removal state $|\Phi^{N-1}_{\mathbf{R}_{I}m\sigma}\rangle$ and $|\Phi^{N-1}_{\mathbf{0}m'\sigma}\rangle$, constructing thereby wave functions $|\breve{\Phi}^{N-1}_{\mathbf{R}_{I}m\sigma}\rangle$ and $|\breve{\Phi}^{N-1}_{\mathbf{0}m'\sigma}\rangle$.
The results indicate small changes of 0.01-0.04 eV for the NN effective hopping terms when relaxation effects are accounted for. The general trend is a slight increase in the NN and NNN hoppings t$_{mm'}$. In each case, the relaxation of the orbital hosting the hole contributes less than 0.01 eV to the overall change of t$_{mm'}$. 
The small increase of the effective hopping terms is attributed to the larger intersite overlap between the relaxed orbitals of the two wave functions$|\breve{\Phi}^{N-1}_{\mathbf{R}_{I}m\sigma'}\rangle$ and $|\breve{\Phi}^{N-1}_{\mathbf{0}m'\sigma'}\rangle$:  the Zn and S valence orbitals within the NN region of the S site hosting the hole are polarized toward the positive charge at this site which leads to a larger intersite orbital overlap.
Analogous trend is observed for the valence-band N $2s$, $2p$ hole states in the more covalent semiconductor zinc-blende BN \cite{AlexJCP}.  
Because of the rather localized nature of the valence-band hole states, the effective hopping matrix elements decay fast with the distance $\mathbf{R}_{I}$. We found the NNN hopping terms to be an order of magnitude smaller than the NN matrix elements, see Table \ref{Hop1}. The small variations in the effective hopping integrals due to short-range relaxation effects lead to a small broadening of the valence S $3s$, $3p$ bands.

To determine the role of
correlation effects beyond relaxation and polarization on the hopping matrix elements, we performed exploratory FO-SDCI calculations. 
As discussed in Section II, the single-particle excitations around the WO accommodating the hole describe short-range relaxation and polarization effects, whereas the two-particle excitations account for correlation effects beyond relaxation and polarization. 
As for the case of diagonal matrix elements, we constructed CI wave functions with single and double excitations from the $3s$ and $3p$ WO's of the active S sites. 
We found that the FO-SDCI treatment leads only to small variations of the NN effective valence-band hoppings of about 0.01 eV. A similar trend was observed for the valence-band N $2s$, $2p$ hole states in zinc-blende BN \cite{AlexJCP}. 
The effect of long-range polarization on the off-diagonal matrix elements is not explicitly considered. This effect, however, is expected to be negligible.

The semicore Zn $3d$ bands exhibit very small dispersion, as inferred from both experimental \cite{dpos1,Langer} and theoretical results \cite{Miyake, Kotani1, Luo, Jaffe}.
Our \textit{ab inito} results show that the localized $3d$ ($N\!-\!1\!$) states are stabilized by on-site 
and short-range relaxation and polarization, see Sec IV. B. 1. 
Since the local relaxation contributions prevail and the delocalization effects are very small, we do not consider explicitly hopping terms associated with the $3d$ hole states. Test calculations for the $3d_{t_2}$ states show that those terms are indeed very small, about 0.01 eV, and the corrections due to short-range relaxation and polarization are even an order of magnitude smaller. Those quantities are expected to be somewhat smaller for the $3d_{e}$ states.

We carried out a similar analysis for the lower-lying Zn $4s$ and $4p$ electron-addtion states and conduction-band NN hopping integrals. We employed thereby the [Zn$_{20}$S$_{40}$] cluster with the $\mathcal{C}_A$ region [Zn$_{2}$S$_{7}$] designed in perfect analogy with that of the [S$_{20}$Zn$_{7}$] cluster. The $\mathcal{C}_B$ region contains all Zn and S ions in the first two coordination shells of the active S sites. We used the 86-311G basis for the S ions and the 86-4111d41G basis for the Zn ions. 

Results from both FOCI and NOCI calculations are summarized in Table \ref{Hop2}. 
In the NOCI calculations, the valence orbitals of the two active Zn ions and the orbitals of their nearest S ions in $\mathcal{C}_A$ are allowed to relax in response to the extra electron in the conduction-band Zn $4s$ or $4p$ orbital.
\begin {table}
 \centering 
 \caption{Nearest-neighbor (NN), $\mathbf{R}_{I}=(0, 1, -1)\frac{a}{2}$ effective hopping matrix elements for the conduction-band states (in eV). FOCI results are listed in the second column. The third column summarizes the NOCI data. Matrix elements smaller than 0.01 eV are not included in the table.}
\begin{tabular}{p{2.8cm}p{2.3cm}p{2.3cm}}
  \hline\hline
Active WO's &\multicolumn{2}{c}{NN t$_{nn'}$($\mathbf{R}_{I}$)} \\ 
&FOCI&RO-NOCI\\ \hline
$4s$-$4s$ &0.081 &0.092 \\
$4p$$_x$-$4p$$_x$ &1.048 & 1.059\\
$4p$$_{y (z)}$-$4p$$_{y (z)}$&0.260& 0.255 \\
$4p$$_{x (y)}$-$4p$$_{y (x)}$& 0.185&0.169 \\
$4p$$_{x (z)}$-$4p$$_{z (x)}$& 0.185&0.169\\
\hline \hline
\end{tabular} 
\label{Hop2}
\end {table}
Compared to the valence-band hoppings, the larger extent of the conduction-band WO's determines larger values of the NN effective hopping terms. 
The correlation-induced corrections due to orbital relaxation and short-range polarization effects, though, are relatively small, i.e., about 0.01 eV. Variations of the hoppings occur in both directions: some matrix elements are slightly increased due to short-range relaxation and polarization, others are reduced.   
Analogous findings were reported for the effective hopping integrals related to the lower-lying conduction-band states in MgO \cite{Hozoi} and BN \cite{AlexJCP}.
Extra effects arising from the relaxation of the orbital hosting the added electron are difficult to compute explicitly because the orbital could not be restrained from delocalizing over other cluster sites in the SCF orbital relaxation. Such an effect, however, is expected to be of the same order of magnitude, 0.01 eV, as that obtained for the ionized S $3s$ and $3p$ states. 

Next-nearest neighbour effective hopping terms are not explicitly considered since such calculations would require much too large clusters, e.g.,  [Zn$_{26}$S$_{50}$], and are thus infeasible with the basis sets required for an accurate study of correlation-induced corrections to electron bands. 
In analogy with the NNN hopping integrals deduced for the B $2s$, $2p$ conduction-band states in BN \cite{AlexJCP}, because of the slower decay of t$_{nn'}$ with the distance $\mathbf{R}_{I}$, some of the NNN t$_{nn'}$ for the Zn $4s$, $4p$ states are expected to be of the same order of magnitude as the NN  t$_{nn'}$.

Basis set effects were assessed by extra calculations at the frozen-orbital HF level for the NN hopping terms, using either the 86-311G basis set for the S ions and the 86-4111d41G basis for the Zn ions or the 86-311G* basis set for S and the 86-4111d311G basis for Zn. 
The cluster employed in these calculations is the [Zn$_{20}$S$_{7}$] cluster. With the smaller basis sets, we found changes of about 0.01 eV in the magnitude of the NN t$_{nn'}$, with the largest deviation of 0.04 eV occurring for the NN Zn $4p_x$-$4p_x$ hopping term. For the NNN conduction-band hoppings, we expect basis set effects of similar magnitude, not larger than few tens of meV.

To summarize, the overall corrections for the intersite interactions due to the short-range polarization cloud around the added electron are found to be small, of the order of 0.01 eV. This outcome is similar to our finding for the $s$ and $p$-derived conduction bands in MgO \cite{Hozoi} and BN \cite{AlexJCP}. 
Hence, the widths of the HF valence and lower-lying Zn $4s$, $4p$ conduction bands change little by incorporating short-range relaxation and polarization effects.

Finally, we consider the effect of the correlation-induced corrections to the widths of the bands on the size of the band gap. We proceed by evaluating the energy $\epsilon_{\mathbf{k}\mu}$ of the S $3p$ and Zn $4s$, $4p$ band states at the $\Gamma_{15}^{v}$ point. The dispersion of the $p_x$ band, for example, reads $\epsilon_{\mathbf{0}\mu, x}$=$E^{(000)}_{x,x}+8t^{(110)}_{x,x}+4t^{(011)}_{x,x}+ O (NNN) + ...$ \cite{Hozoi, AlexJCP}, where $E^{(000)}_{x,x}$ is a diagonal matrix element and $t^{(110)}_{x,x}=t^{(011)}_{y,y}=t^{(011)}_{z,z}$ by symmetry. 
Corrections of about 0.03 eV for $t^{(110)}_{x,x}$ and 0.005 eV for $t^{(011)}_{x,x}$ (see Table \ref{Hop1}) lead to an upward shift of the S $3p$ bands by 0.196 eV at the $\Gamma$ point. 
The band gap is reduced then by the same amount.   

We have not computed explicitly similar corrections for the Zn $3d$ states. Such corrections are expected to shift the Zn $3d$ bands upwards by little, such that the average energy position $E_d$ of the Zn $3d$ states relative to the top of the valence S $3p$ bands at the $\Gamma$ point will remain about --9.0 eV.

The correlation-induced corrections to the Zn $4p$-$4p$ and Zn $4s$-$4s$ intersite matrix elements are relevant at the $\Gamma_{1}^{c}$ symmetry point. The energy of the $4s$ band at $\Gamma_{1}^{c}$ is given by $\epsilon_{\mathbf{k}\nu, s}$=$E^{(000)}_{s,s}$+12t$^{(110)}_{s,s}$+... \cite{SlaterKoster}. For symmetry reasons, t$^{(110)}_{s,s}$=t$^{(011)}_{s,s}$. Corrections of 0.011 eV for t$^{(011)}_{s,s}$ lead to a shift to lower energy of the Zn $4s$ band at $\Gamma_{1}^{c}$ and a further reduction of the gap by 0.132 eV. The Zn $4p$ bands are also shifted downwards by about 0.08 eV. 

The correlation-induced corrections to the off-diagonal conduction-band matrix elements further reduce thus the size of the gap to 5.16 eV. 
This value is still off by 1.36 eV as compared to the experiment. 
We have excluded from the current investigation differential correlation effects beyond the loss of ground-state correlations. The overall contribution of those effects to the correlation-induced corrections to the valence-band Zn $3d$, S $3p$ and conduction-band Zn $4s$, $4p$ states is expected to lead to a further reduction of the gap. 
Studies of correlation effects beyond the loss of ground-state correlations are left for future considerations. 
Further, we have not explicitly considered the S $4s$ and $4p$ conduction-band states. For those states, delocalization effects are expected to dominate over local relaxation and polarization. As discussed in Section IV. A, the WO associated with the dispersive conduction band of predominantly S $4s$, $4p$ character which is entangled with the lower-lying Zn $4s$, $4p$ conduction bands is spread over four S ions and hence, very large clusters are needed to describe such a ($N\!+\!1\!$) state. 

\section {Conclusions}

We have presented a detailed study of the valence- and low-lying
conduction bands of zinc-blende $c$-ZnS. We have put thereby special
emphasis on the ``semicore'' Zn $3d$ band states. The starting point in the study was a HF
bandstructure. Correlation-induced corrections to it were computed based
on a local Hamiltonian approach and a quasiparticle approximation. The 
aim of our investigation was to demonstrate that wave-function-based methods are able to deal
in an \textit{ab initio} manner with the topic of band structures without
compromising on the infinite range of the correlation holes. This is
important here, since an added electron (conduction band) or hole
(valence band) generates a long-range polarization cloud around itself.
Long-range polarization effects, which add to short-range
relaxation and polarization, contribute to the reduction of the HF gap
by about 1.5 eV, and therefore are not negligible.

The new aspect in this study as compared with previous calculations on MgO and BN is
the inclusion in the treatment of the aforementioned $3d$ semicore states with which
density functional based calculations have problems. Overall
correlation-induced corrections to the valence-band energies place the
Zn $3d$ semicore states at about --9.0 eV below the top of the S $3p$ valence
bands, in agreement with x-ray photoemission spectroscopy \cite{dpos1}. Summing up
all correlation-induced corrections, the size of the band gap reduces to
5.5 eV as compared with the HF gap of 11.0 eV. An additional, though
small, reduction of the gap by 0.3 eV originates from correlation-induced
variations in the widths of the HF bands. Correlation effects beyond the computed loss of
ground-state correlation, when an electron or hole is added, are
expected to decrease further the value for the gap and bring it closer
to the experimental estimate of 3.8 eV.

In summary, we may state that wave-function-based calculations of
energy bands as performed in the present study provide a promising approach, which is free of
any uncontrolled approximations and arbitrariness. They deserve continuing attention.

\appendix
\section{86-4111d311G basis sets for Zn.}
\begin {table}[h]
 \centering 
 \caption{Exponents (in a.u.) and coefficients of the Zn 86-4111d311G GTO basis set.}
\begin{tabular}{p{1.2cm}p{1.3cm}p{1.9cm}p{1.9cm}p{2.0cm}}
  \hline\hline
& & &\multicolumn{2}{c}{Coefficient} \\ 
Shell&Type &Exponent & s&p\\ \hline 
4&sp & 1.68680&1.0&1.0 \\
5&sp &0.62679&1.0&1.0\\
6&sp &0.15333 & 1.0&1.0\\ \hline
\end{tabular}
\begin{tabular}{p{1.2cm}p{1.3cm}p{2.9cm}p{3.0cm}}\hline
& & &\multicolumn{1}{c}{Coefficient} \\
Shell&Type &Exponent & d\\ \hline 
3&d&56.079664803 &  0.029598233794 \\
&&15.748259098 &  0.15879184876 \\
&&5.3098927689 &  0.37984178082 \\
4&d&1.7734652510 &  1.0\\
5&d&0.51963233927&1.0\\ \hline\hline
\end{tabular} 
\label{BS1}
\end {table}
\newpage

\section{Detailed account of the loss of ground-state correlations.}

\paragraph {Zn $3d$ ($N\!-\!1\!$) states.}
As discussed in Sec. IV. B. 5, the SDCI wave functions for the $N$-electron GS and Zn $3d_{t_2}$ or $3d_{e}$ hole states are constructed by correlating the $3s$, $3p$, and $3d$ orbitals of the Zn$^{c}$ site in the $\mathcal{C}_A$ region [Zn$^{c}$S$_{4}$Zn$_{12}$] plus the $3s$ and $3p$ orbitals of the four NN S ions. 
The reference wave functions $|\breve{\Phi}^{N-1}_{\mathbf{R}_{I}m\sigma}\rangle$ are each expressed in terms of individually optimized orbital sets.    
All singlet and triplet electron pairs formed by electrons from the doubly occupied Zn $3s$, $3p$, $3d$ and S $3s$, $3p$ orbitals and from the singly occupied Zn $3d_{t_2}$ or $3d_{e}$ orbital are included in the correlation treatment. 

The SDCI correlation energy for the $N$-electron state is found to be --22.56 eV. This correlation energy is partitioned into contributions of excited internal, singly external, and doubly external configurations. The notation for the different configurations follows that by Werner and Knowles in Ref.\ ~\onlinecite{ICCI}.
The excited singly external configurations arise from one-particle and semi-internal two-particle excitations from occupied to virtual (external) orbitals. 
The semi-internal two-particle excitations are two-particle excitations under the constraint that exactly one electron is excited from a singly or doubly occupied orbital to an external orbital. The excited doubly external configurations result from external two-particle excitations where both electrons are promoted to virtual orbitals. 
The external two-particle excitations contribute with --22.53 eV to the total correlation energy for the $N$-electron  state. The very small contribution of --0.03 eV arising from singly external configurations is associated with the self-consistency of the cluster HF wave function. 
The coefficient of the reference RHF wave function in the SDCI calculation is 0.92. 

The SDCI correlation energies for the Zn $3d_{t_2}$ and $3d_{e}$ electron-removal states amount to --21.54 eV and --21.43 eV, respectively. 
The coefficients of the reference ROHF wave functions in the SDCI  
calculations are 0.92. 
In analogy to the $N$-electron state, the major contributions result from external two-particle excitations, --20.93 and --21.07 eV, respectively. For the Zn $3d$ ($N\!-\!1\!$) states, the contribution due to singly external configurations is moderate, --0.26 eV.
With the present choice of the cluster, the correlation contribution of excited internal configurations is about --0.09 eV for the $3d_{e}$ ($N\!-\!1\!$) states and about --0.34 eV for the $3d_{t_2}$ hole states. 
The excited internal configurations with the Zn $3d_{t_2}$ or $3d_{e}$ orbital doubly occupied are mainly related to excitations from the doubly occupied S $3s$, $3p$ or central Zn $3p$ orbitals.
The smaller magnitude of this contribution for the $3d_{e}$ ($N\!-\!1\!$) states is due to the orientation of the $3d_{e}$ orbitals with respect to the charge distributions related to the bonding and non-bonding electron pairs at the four S sites in [ZnS$_{4}$] . 

In the framework of the quasiparticle picture and frozen local hole approximation \cite{Pahl}, the occupation of the spin-orbital hosting the hole should be kept frozen in all configurational state functions (CSF's) in the SDCI expansions.     
We work, however, with spatial orbital rather than spin-orbital sets, and the configuration selection in \textsc{molpro} is done for spatial orbitals. 
Therefore, the SDCI wave functions constructed by correlating the singly occupied hole orbital and the doubly occupied orbitals contain CSF's where the occupancy of the hole orbital is no longer maintained. 
The contributions arising from such CSF's to the SDCI correlation energies are related to effects beyond the loss of ground-state correlation and the quasiparticle picture. 
In the calculation of the loss of ground-state correlations such contributions should be thus disregarded. These additional differential correlation effects may be needed, for example, for describing satellite structures in the photoionization spectra.

With the present variational orbital space for the $3d$ ($N\!-\!1\!$) states, the occupation of the orbital hosting the hole remains unaltered in the excited doubly external configurations. Hence, the correlation contributions of those configurations are relevant in the calculation of the loss of ground-state correlations. Contributions arising from excited internal and singly external configurations where the occupation of the hole orbital is altered are on the other hand disregarded. The energy contributions of excited internal configurations, --0.34 eV and --0.09 eV, are thus excluded. 
The correlation contribution of --0.26 eV associated with the singly external configurations is 
further analyzed in a series of SDCI calculations in which only certain orbitals or groups of orbitals (e.g., pairs, triples) are explicitly correlated. We found that a major fraction of that contribution is obtained when only the Zn $3s$, $3p$ and $3d$ orbitals are explicitly correlated, --0.21 eV for the $3d$ ($N\!-\!1\!$) states. When the Zn $3s$, $3p$ orbitals alone are correlated, the contribution of singly external configurations is found to be --0.11 eV. This correlation energy is, however, discarded because of the significant presence in the SDCI expansion of excited singly external CSF's with both $3d_{t_2}$ and $3d_{e}$ orbitals doubly occupied.
Correlating explicitly only the Zn $3d$ orbitals results into an energy contribution of --0.07 eV as regards the singly external configurations. This contribution is also excluded from the computation of the loss of ground-state correlations.
Finally, correlating the relaxed S $3s$ and $3p$ orbitals alone leads to a small contribution of --0.03 eV, which is also disregarded because of the altered occupation of the hole orbital in the singly external configurations. The correlation contribution of --0.26 eV is thus mainly related to configurations with an altered occupation of the $3d$ hole orbitals.
Therefore, the loss of ground-state correlations may be deduced by comparing only the correlation contributions due to excited doubly external configurations for the $3d$ ($N\!-\!1\!$) states and for the $N$-electron ground state. 
The corrections to the diagonal matrix elements due to this effect shift the Zn $3d_{t_2}$ and $3d_{e}$ bands downward by 1.60 and 1.46 eV, respectively (see Table \ref{diagonalhole}). 

\paragraph {S $3s$ and $3p$ ($N\!-\!1\!$) states.} 
SDCI wave functions for the $N$-electron and S $3s$ and $3p$ ($N\!-\!1\!$) states are constructed by correlating the S $3s$, $3p$ and Zn $3d$ orbitals in the active region [SZn$_{4}$] of the [S$_{13}$Zn$_{28}$] cluster. 
The SDCI correlation energy for the $N$-electron ground state is found to be --24.16 eV with the largest contribution of --24.12 eV arising from excited doubly external configurations.

For the S $3p$ ($N\!-\!1\!$) states, the SDCI energy amounts to --23.98 eV, with a dominant contribution of --23.30 eV arising from excited doubly external configurations. The largest fraction of the latter contribution, --20.93 eV, is due to correlation effects within the Zn $3d$ shell, as inferred from a separate  SDCI calculation where only the Zn $3d$ doubly occupied orbitals were explicitly correlated. 
Excited internal configurations give rise to a small correlation contribution of --0.13 eV and imply charge transfer from the Zn $3d$ doubly occupied orbitals, delocalized over the four Zn sites, to the S $3p$ hole orbital.   
A moderate contribution of --0.54 eV arises from 
excited singly external configurations. 
Correlating the Zn $3d$ doubly occupied orbitals alone brings that contribution to --0.07 eV. The major fraction of those --0.54 eV originates thus from correlation effects within the S $3s$ and $3p$ orbital space. 
This finding is also supported by separate SDCI calculations where only the S $3s$, $3p$ and core-like Zn $3p$ orbitals are explicitly correlated. 
With the present cluster and such a variational orbital space, the singly external configurations are found to arise from excitations involving only the S $3s$ and $3p$ orbitals, with an overall energy effect of --0.53 eV. 
An inspection of the corresponding SDCI expansion indicates considerable presence of CSF's with altered occupation of the S $3p$ hole orbital as compared to that in the reference ROHF wave function. The correlation contribution associated with the singly external configurations, --0.54 eV, has been thus discarded from the computation of loss of ground-state correlations.  
To deduce the loss of ground-state correlations we utilized for the $N$ and S $3p$ ($N\!-\!1\!$) states the dominant correlation contributions due to doubly external configurations. We found a correction of 0.82 eV to the S $3p$ band states, see Table \ref{diagonalhole}. 

The loss of ground-state correlations for the S $3s$ hole state is difficult to compute because maintaining the hole in the S $3s$ orbital in all configurations of the SDCI wave function is technically impossible. 
We calculated the associated correction by using only the correlation contribution of the excited doubly external configurations in the wave function $|\Psi^{N-1}_{\mathbf{R}_{I}m\sigma}\rangle$ of the S $3s$ hole state. 
The correlation contribution resulting from excited singly external configurations is largely due to semi-internal two-particle excitations which involve the S $3s$ and S $3p$ orbitals and lead to a doubly occupied S $3s$ orbital. 
It amounts to about --2.5 eV.
With the present choice of the finite cluster and variational orbital space, the internal configurations contribute with --0.42 eV to the SDCI energy. They are related to Zn $3d$ to S $3s$ charge transfer, just as in the case of the S $3p$ ($N\!-\!1\!$) states. 
The overall correlation correction obtained with the selected variational space shifts downward the center of gravity of the S $3s$ bands by 1.76 eV. 

\paragraph{Zn $4s$, $4p$ ($N\!+\!1\!$) states.}
Analogous considerations were applied to the low-lying conduction-band Zn $4s$, $4p$ states.
A [Zn$_{13}$S$_{28}$] cluster with the active region [Zn$^{c}$S$_{4}$Zn$_{12}$] was used for the study of the ($N\!+\!1\!$) states and the S 86-311G* and Zn 86-4111d311G basis sets. 
We constructed the SDCI wave functions for the $N$ and ($N\!+\!1\!$) states by correlating explicitly the doubly occupied  $3s$,  $3p$, and $3d$ orbitals of Zn$^{c}$ and the doubly occupied $3s$ and $3p$ orbitals of the four NN S ions. For the ($N\!+\!1\!$) states we  
started from the wave functions $|\tilde{\Phi}^{N+1}_{\mathbf{R}_{I}n\sigma}\rangle$ where the orbitals at the central Zn$^{c}$ ion and those at the four S ions around Zn$^{c}$ were previously relaxed. 
The singly occupied Zn $4s$-like or $4p$-like orbital is placed into the active orbital space but no excitations out of this orbital are included in the correlation treatment. 
The SDCI wave function constructed in this manner incorporates correlation contributions of external one- and two-particle excitations involving electrons or electron pairs from the doubly occupied orbitals as well as contributions of internal excitations and semi-internal two-particle excitations for which an electron is promoted from a doubly occupied orbital to the Zn $4s$ or $4p$ orbital. 

To evaluate the loss of ground-state correlations, we used the total SDCI correlation energies obtained for the $N$ and  ($N\!+\!1\!$) states. 
The corrections to the diagonal matrix elements $\Delta H^{lgsc}_{nn}(\mathbf{0})$ due to the loss of ground-state correlations are about the same for the  Zn $4s$ and $4p$ electron-addition states, 0.1 eV.

SDCI wave functions constructed by including in the correlation treatment not only the doubly occupied orbitals but also the singly occupied Zn $4s$-like or $4p$-like orbital account for correlation effects beyond the loss of ground-state correlations. Such correlation effects are responsible, for example, for the presence of satellite structures in the inverse photoionization spectra. 
The inclusion of the singly occupied Zn $4s$-like orbital 
in the correlation calculation results, for example, into an extra differential contribution of --0.97 eV to the afore-discussed SDCI energy of the Zn $4s$ ($N\!+\!1\!$) state. 
This extra contribution is mainly due to excited doubly external configurations, about --0.70 eV, but singly external configurations involving the single electron give rise to a relevant fraction of about --0.27 eV.  
We note that the virtual orbitals to which the extra electron is excited in the singly external configurations with largest CI coefficients have much weight at the four NN S ions.
This extra differential contribution of --0.97 eV stabilizes the ($N\!+\!1\!$) electron state with respect to the $N$-electron state by --0.84 eV. 
It includes, however, a fraction due to a delocalization SCF effect since some of the projected virtual orbitals in the variational orbital space to which the Zn $4s$ electron is excited have also weight at the twelve Zn sites in the [Zn$^{c}$S$_{4}$Zn$_{12}$] region. 
We have not computed explicitly the magnitude of the SCF delocalization effect. Our results for the NN Zn $4s$-$4s$ effective hopping matrix elements at the frozen-orbital HF level (see Table \ref{Hop2}) indicate that the delocalization of the extra $4s$ electron over the twelve NN Zn sites is associated with a stabilization energy of about 12t$^{(011)}_{s,s}$=--0.972 eV.

\end{document}